\documentclass[12pt]{article}

\usepackage{jcappub}
\usepackage{color}
\usepackage{subfigure}

\usepackage{graphics,graphicx}
\usepackage{bm}
\usepackage{amsmath,amssymb}
\usepackage{dcolumn}
\usepackage{multirow}
\usepackage{fancyhdr}

\usepackage{arydshln}
\usepackage{booktabs}
\usepackage{graphicx}
\usepackage{url}
\usepackage{amsbsy} 

\newcommand{\physrep}{Phys. Rep.}

\newcommand{\apjl}{Astrophys. J Lett.}

\newcommand{\apj}{Astrophys. J}

\newcommand{\prd}{Phys. Rev. D}

\newcommand{\mnras}{Mon. Not. R. Astron. Soc.}

\newcommand{\nat}{Nature}
\newcommand{\jcap}{JCAP}

\def\gsim{\;\rlap{\lower 2.5pt
 \hbox{$\sim$}}\raise 1.5pt\hbox{$>$}\;}
\def\lsim{\;\rlap{\lower 2.5pt
 \hbox{$\sim$}}\raise 1.5pt\hbox{$<$}\;}

\def\cl{\makebox[10mm][l]{CMB ($T$,$E$,$\psi$)}}
\def\w{$\gamma$}
\def\g{${\rm g}$}
\def\clw{CMB\,+\,$\gamma$}
\def\clg{CMB\,+\,g}
\def\clgw{\makebox[10mm][l]{CMB\,+\,g\,+\,$\gamma$}}
\def\clgwsm{\multicolumn{14}{l||}{{CMB\,+\,g\,+\,$\gamma$}}}
\def\clgws{\hspace{0.5pt} +\,$g^{\rm 3D}$(spec-$z$)}

\def\be{\begin{equation}}
\def\ee{\end{equation}}
\def\ba{\begin{eqnarray}}
\def\ea{\end{eqnarray}}
\def\bv{\begin{verbatim}}
\def\ev{\end{verbatim}}
\def\nn{\nonumber}

\def\bea{\begin{eqnarray}}
\def\eea{\end{eqnarray}}

\def\Mpc{{\rm Mpc}}
\def\Gpc{{\rm Gpc}}

\def\lmax{\ell_{\rm max}}

\def\kmax{k_{\rm max}}
\def\knl{k_{\rm nl}}

\def\nh{n_{\rm h}}

\def\Mmin{M_{\rm min}}

\def\Msun{M_{\odot}}

\def\fsky{f_{\rm sky}}

\def\bng{\bar{n}_{\rm g}}

\def\zc{z_{\rm c}}
\def\zm{z_{\rm m}}
\def\Vs{V_{\rm s}}

\def\Msun{M_\odot}

\def\Mpc{{\rm Mpc}}
\def\Gpc{{\rm Gpc}}

\def\Yp{Y_{\rm p}}
\def\Neff{N_{\rm eff}}
\def\nmsum{\sum m_{\nu}}

\title{
Probing Neutrinos from Planck and Forthcoming Galaxy Redshift Surveys
}

\author{Yoshitaka Takeuchi and Kenji Kadota}

\affiliation{Department of Physics, Nagoya University, Nagoya 464-8602, Japan}

\emailAdd{yoshitaka@nagoya-u.jp}
\emailAdd{kadota.kenji@f.nagoya-u.jp}

\abstract{

We investigate how much the constraints on the neutrino properties can be improved
by combining the CMB, the photometric and spectroscopic galaxy
redshift surveys which include the CMB lensing, galaxy lensing
tomography, galaxy clustering  and redshift space distortion observables.  We pay a particular attention to the constraint on
the neutrino mass in view of the forthcoming redshift surveys such as
the Euclid satellite and the LSST survey along with the Planck CMB lensing measurements. 
Combining the transverse mode information from the angular power
spectrum and the longitudinal mode information from the spectroscopic survey with the redshift
space distortion measurements can determine the total neutrino mass
with the projected error of ${\cal O}(0.02)$eV. 
Our
analysis fixes the mass splittings among the neutrino species to be
consistent with the neutrino oscillation data, and we accordingly study the
sensitivity of our parameter estimations on the minimal neutrino mass. The
cosmological measurement of the total neutrino mass can distinguish
between the normal and inverted mass hierarchy scenarios if the minimal
neutrino mass $\lesssim 0.005$ eV with the predicted 1-$\sigma$
uncertainties taken into account.

}


\begin{document}

\maketitle


\section{Introduction}
\label{set:intro}

There has been an ever-growing evidence supporting the non-vanishing
neutrino masses from the measurements of the neutrino oscillations
\cite{pmns1,pmns2} including the solar ($\nu_e$), atmospheric
($\nu_{\mu}, \bar{\nu}_{\mu}$), reactor ($\bar{\nu}_e$) and
accelerator $(\nu_{\mu})$ neutrino experiments \cite{pdg}.
There also have been the precise measurements of the cosmological
observables such as the cosmic microwave background (CMB) which favor
the non-vanishing neutrino masses \cite{spt2,plsz}.

We in this paper forecast the strong constraints on the neutrino
masses from the observations of the large-scale structure (LSS) and
CMB.  The different astrophysical probes on the neutrino masses are of
great interest for their complimentarity and for their reducing the
degeneracies among the relevant parameters.
The current cosmological data have put the upper bounds of order
$\nmsum \lesssim 0.2$-0.5 eV and there also have been numerous
theoretical studies for the future forecasts on the neutrino mass
constraints claiming even an order of magnitude improvements from the
cosmological measurements such as the CMB, galaxy weak lensing, galaxy
clustering, galaxy cluster count, baryon acoustic oscillations, Type
Ia supernovae and Lyman-$\alpha$ forest 
\cite{
  Perotto:2006,Galli:2010,Audren:2013,
  kev,uros,sanc,mant,bure,
  debe,scott02,berna,ichi,hanne,vik,acq,car,hama,cer,yun,van,manoj,wan,les2,hall12,namikawa,takada,hudo,fogl,gonza,giusa}. 
For the CMB and LSS observables, for instance, the unprecedented
precision on the neutrino mass constraints can be expected from the
on-going and forthcoming experiments such as the Planck \cite{pl}, SPTPol \cite{sptpol},
ACTPol \cite{actpol} providing the CMB lensing measurements, and a number of wide-field surveys for both
ground-based imaging surveys and space-based missions such as 
the Subaru Hyper Suprime-Cam Survey (HSC\footnote{http://www.naoj.org/Projects/HSC/index.html}) \cite{HSC}, 
the Dark Energy Survey (DES\footnote{http://darkenergysurvey.org}) \cite{DES}, 
the Large Synoptic Sky Survey (LSST\footnote{http://www.lsst.org}) \cite{LSST}, 
and 
the European Space Agency (ESA) Euclid satellite mission\footnote{http://www.euclid-ec.org} \cite{Euclid}.

To realize the tighter constraints on the neutrinos, however, we need
to break the degeneracies among the cosmological parameters, e.g. the
degeneracies among the neutrino mass and the dark energy parameters.
One of the sound tactics to lift the degeneracies is to combine
multiple observables and to take the cross-correlations among those
observables which have the different dependence on the cosmological
parameters of our interest.  Our study therefore emphasizes the merit
of combining the angular power spectrum, in particular the CMB lensing
and photometric galaxy lensing tomography, and the three-dimensional
galaxy power spectrum from the spectroscopic redshift survey which
possesses not only the transverse mode information but also the
information along the line of sight.
We, for this purpose, perform a Fisher matrix analysis by combining
the angular power spectrum and the three-dimensional galaxy power
spectrum, and we investigate how accurately the property of the
neutrino can be constrained from the on-going and upcoming surveys.
We, for concreteness, adopt the Planck, LSST and Euclid surveys as the
reference surveys in the following, deferring to the discussion
section the quantitative comparison with the analogous analysis using
the on-going experiments HSC and DES instead of the future experiment
LSST.

We also seek the distinction between two neutrino mass structures,
normal $m_{\nu_1}<m_{\nu_2}\ll m_{\nu_3}$ and inverted $m_{\nu_3}\ll
m_{\nu_1}< m_{\nu_2}$ mass hierarchy scenarios which are the fiducial
models in our forecasts. For the fiducial values of the
neutrino mass splittings, we take the best fit values for the
atmospheric neutrino $|\Delta m_{31(32)}^2|$ (the parenthesis is for
the inverted hierarchy) and the solar neutrino parameter $\Delta m_{21}^2$
based on the global analysis of neutrino oscillation data including
the recent measurements of the neutrino mixing angle $\theta_{13}$ at
the reactor experiments \cite{fogi3} 
\ba
\label{ndata}
  |\Delta m_{31(32)}^2|=2.47 (2.46) \times 10^{-3}~{\rm eV}^2 ~ , \hspace{5mm}
  \Delta m_{21}^2=7.54 \times 10^{-5}~{\rm eV}^2 ~ , 
\label{eq:dmv}
\ea
where $\Delta m_{ij} \equiv m_{\nu_i}^2-m_{\nu_j}^2 $. The minimal neutrino mass
$m_{\nu,{\rm min}}$, which corresponds to $m_{\nu_1}(m_{\nu_3})$ for the normal (inverted) mass
spectrum, is undetermined from the oscillation data, and we also study the
sensitivity of the neutrino parameter estimation on the value of $m_{\nu,{\rm min}}$.

This paper is organized as follows. We first briefly review the
effects of the massive neutrinos on the cosmological observables in
Section~\ref{sec:neu}. In Section~\ref{sec:obs}, we summarize the
cosmological observables used in the analysis and how those
observables are theoretically treated in terms of the angular power
spectrum and the three-dimensional galaxy power spectrum.
Section~\ref{sec:fisher} describes our Fisher matrix analysis and
Section~\ref{sec:result} presents the main results on the improvement
in the neutrino property constraints by combining the CMB, the
photometric redshift survey and the spectroscopic redshift
survey. Section~\ref{sec:discuss} is devoted to the discussion and
conclusion.


\section{The massive neutrinos in cosmology}
\label{sec:neu}
We start with a brief review on the effects of the neutrinos on the
CMB and structure formation.  Because a neutrino becomes
non-relativistic around $1+z\sim 300$ ($m_{\nu}/ 0.05$ eV) when the
temperature becomes comparable to its mass, we assume, unless stated
otherwise, the sub eV mass neutrinos of our interest become
non-relativistic during the matter domination era.

The massive neutrinos lighter than a half eV are relativistic at the
recombination epoch and their effects of reducing the matter-radiation
ratio on the CMB are manifest in the early integrated Sachs-Wolfe (ISW) effect. The massive
neutrinos deep in the matter domination epoch and in the dark energy
domination epoch can also cause the decay of the gravitational
potential enhancing the late ISW effect
\cite{uros2,rep,scott2,les2,hudo}. 
Those effects imprinted in the primary CMB anisotropy can however be
also induced by varying other cosmological quantities such as 
the Hubble parameter and the equation of state parameter $w$
\cite{dick3,kev2,how}. The CMB lensing turns out to be quite powerful
in breaking the parameter degeneracies which would otherwise be
persistent in the CMB power spectrum because the lensing is sensitive
not only to the geometry but also to the growth of the structure along
the line of sight, and the recent impressive progress on the CMB
lensing measurements would be of great interest for the exploration of the neutrino properties. The
CMB lensing reconstruction from the temperature anisotropies alone has
the limitation due to its statistical noise and it benefits
significantly from the polarization measurements probing the smaller
scales with high precision \cite{take}.

The matter power spectrum can be affected by the neutrinos because the structure formation is suppressed
due to the free-streaming of the neutrinos for the scales below the
neutrino Jeans length scale \cite{bond}. The neutrino comoving
free-streaming scale decreases as $\lambda_{\rm FS}\propto a^{-1/2}$
during the matter domination so that the turn-over scale of the
matter power spectrum due to the neutrino suppression has the upper
bound corresponding to the neutrino free-streaming scale when it
becomes non-relativistic. Each different mass neutrino has a different
free-streaming scale and hence the matter power spectrum in principle
can probe each neutrino mass rather than just a sum of neutrino
masses. For instance, even if the value for a total neutrino mass is
identical, the error on the total neutrino mass from a galaxy redshift
survey can differ depending on the neutrino mass splitting patterns
\cite{les04,takada,yun}.  The neutrino free-streaming scale during
the matter domination is 
\ba
  k_{\rm FS}\sim 0.015 {\rm Mpc}^{-1} \left(\frac{m_{\nu}}{0.05 {\rm eV}} \right) \sqrt{\frac{\Omega_{\rm m} h^2}{0.14}\frac{1}{1+z}} 
\ea
This suppression turn-over scale is proportional to the neutrino mass,
hence the finding this suppression scale can probe each neutrino mass
scale while the overall amplitude suppression gives us the information
on the sum of the neutrino masses. Too light a neutrino becomes hard
to be probed by the galaxy survey because $k_{\rm FS}$ may become too
small for the survey volume to cover, even though other measurements
such as the CMB observables can compliment such deficiencies. The
free-streaming scale is also dependent on the redshift, and the
spectroscopic galaxy redshift survey with the accurate redshift
information would be of great help to probe the neutrino features
imprinted in the matter fluctuations.  For $k<k_{\rm FS}$, the
non-relativistic neutrinos falls into the gravitational potential
along with the cold dark matter (CDM) and baryons, and the conventional picture for
the matter fluctuation growth proportional to the standard linear growth
factor $D(z)$ applies. For $k>k_{\rm FS}$ with the neutrino free-streaming
effects operative, the non-relativistic neutrinos cannot cluster
because of too large a velocity dispersion. The growth of the matter
fluctuations hence is slower compared with that for $k<k_{\rm FS}$, and
the suppressed power is proportional to $(1-f_{\nu}) D(z)^{1-p}$ for
$k\gg k_{\rm FS}$ with $p=(5-\sqrt{25-24 f_{\nu}})/4$ and $f_{\nu}\equiv \Omega_{\nu}/\Omega_{\rm m}$
\cite{bond,eh97,eh98,rep}. Hence, for a scale
below the current free-streaming scale, the matter power spectrum is
suppressed roughly by a fraction $\Delta P_{\rm m}(k)/P_{\rm m}(k)
\sim -8 f_{\nu}
$ with an even bigger suppression in the
non-linear regime \cite{hu98b,bir}). 
Due to the dependence of the matter power on $f_{\nu} \propto
\nmsum/\Omega_{\rm m} h^2$, the neutrino mass is also degenerate with $\Omega_{\rm m} h^2$ \cite{hu98b}
even though the precise measurement of $\Omega_{\rm m}$ from other
experiments such as the Planck makes $m_{\nu}$-$\Omega_{\rm m} h^2$
degeneracy less significant. We also benefit from combing the linear
scale galaxy redshift survey and the galaxy lensing survey because the
lensing is disadvantageous for being sensitive to the mass clustering
at the non-linear scales. 
A potential problem in probing the galaxy clustering is the issue of
the bias which hinders the direct probe on the matter clustering and
hence leads to the uncertainties in the neutrino mass
constraints. Combing the spectroscopic redshift survey data and the
galaxy weak lensing tomography data can help reduce the bias
uncertainty 
and hence can improve
the neutrino mass constraints as we shall discuss in the following.


\section{Cosmological observables}
\label{sec:obs}

We here summarize the cosmological observables used in our
analysis. In this work, we investigate the constraints from the data
combining three kinds of cosmological surveys: the CMB experiment, the
photometric redshift survey and the spectroscopic redshift survey. 
 The observables from the
CMB experiment and the photometric redshift survey are the projected
two-dimensional data, providing the angular power spectra
$C_\ell$. The spectroscopic redshift
survey can give us the three-dimensional
power spectrum $P({\bm k})$ providing the information for not only the transverse direction but
also for the line-of-sight direction.

\subsection{Angular power spectrum}
\label{sec:cls}

The angular power spectrum for auto- and cross-correlations of the observables $X$ and $Y$ can be given by 
\ba
C_\ell^{XY} = \frac{2}{\pi}\int k^2 dk P_{\rm m}(k) \Delta_\ell^X(k)\Delta_\ell^Y(k) ,
\ea 
where $P_{\rm m}(k)$ is the matter power spectrum at present and
$\Delta_\ell^X$ is the kernel of the observable $X$. We consider the
CMB temperature anisotropies ($T$), $E$-mode polarization ($E$), CMB
lensing potential ($\psi$), the galaxy distributions (${\rm g}_i$) and
the galaxy weak lensing shear ($\gamma_i$), i.e. $X \in
\{T,E,\psi,g_i,\gamma_i\}$, as the observables for the angular power
spectra. The observables $T$ and $E$ are the unlensed ones, and the
information of the CMB lensing is included just in the CMB lensing
potential $\psi$. The subscript on the galaxy distributions and the
weak lensing shear represents the $i$-th redshift bin in the
tomographic surveys.  \\

The temperature and $E$-mode polarization angular power spectra are
obtained from a publicly available code CAMB \cite{Lewis:2000}. The
cross-correlation for $T{\rm g}_i$ and that for $T\gamma_i$ arise from
the late ISW effect \cite{Sachs:1967} and
the kernel of the late ISW is given by
\ba
  \Delta_\ell^{\rm ISW}(k)=3 \Omega_{\rm m,0} H_0^2 \int _0 ^{z_*} dz \frac{d}{dz}[D(z,k)(1+z)] \frac{j_\ell(kr(z))}{k^2} ~ ,
\ea
where $z_*$ represents the redshift of the last scattering surface,
$r(z)$ is the comoving distance to the redshift $z$, $\Omega_{\rm
  m,0}$ and $H_0$ are respectively the matter density parameter and
the Hubble parameter at the present time. The function $D(z,k)\equiv
\sqrt{P_{\rm m}(k,z)/P_{\rm m}(k,0)}$ represents the growth factor,
$P_{\rm m}(k,z)$ represents the matter power spectrum at redshift $z$
and $j_\ell(kr(z))$ is the spherical Bessel function.

The CMB photons coming from the last scattering surface are deflected
by the potential gradients of the LSS along the line-of-sight
\cite{Lewis:2006}, hence the
CMB lensing potential also has the correlation with the LSS. For
the CMB lensing potential, the kernel is give by
\ba
  \Delta_\ell^{\psi}(k)=3 \Omega_{\rm m,0}H_0^2 \int ^{r_*}_{0}dr 
  \frac{r_*-r}{r r_*}
  \frac{D(z(r),k)}{a(r)} \frac{j_\ell (kr)}{k^2} ,
\ea
where $r_*$ represents the comoving distance to the last scattering
surface. 
\\

We can obtain two kinds of observables from the photometric
redshift survey; one is the distribution of galaxies and the other is
the galaxy weak lensing. For the galaxy distributions, we consider the
tomographic survey with some redshift bins, and the kernel of the
galaxy distributions is given by
\ba 
  \Delta_\ell^{g_i}(k) = 
  \int_{z_i}^{z_{i+1}} dz \; b(z) \frac{n_i(z)}{n_i^{\rm A}} 
  D(z,k) j_\ell(k r(z)) ,
\label{eq:delta_g}
\ea
where $b(z)$ is the biasing parameter of galaxy and 
$n(z)$ and $n_i^{\rm A}$ represent the redshift distribution of sample
galaxies and its normalization factor respectively.  We adopt a
time-varying linear biasing parameter by the first-order expansion
$b(z)=b_0+b_1 z$. We treat these two model parameters $b_0$ and $b_1$
as free parameters in our analysis and the fiducial values are set to
($b_0$,$b_1$)=(1.0,0.8) \cite{Weinberg:2004}.

We also assume the tomographic survey for the galaxy weak lensing. The galaxy weak lensing survey measure the shear of source
galaxies due to the distributions of foreground galaxies. For the galaxy
weak lensing shear, the kernel is given by
\ba
  \Delta_\ell^{\gamma_i}(k)=\int_0^{\infty}  dr 
  W_{\ell,i}(r(z))  D(z(r),k) \frac{j_\ell(kr(z))}{k^2} ~ ,
\label{eq:shear}
\ea
with
\ba
  W_{\ell,i}(r(z))=\frac{3\Omega_{\rm m,0}H_0^2}{2a r(z)} \sqrt{\frac{(\ell+2)!} {(\ell-2)!}} 
  \int_{{\rm max}(z,z_i)}^{z_{i+1}} dz_s \frac{n(z_s)}{n_i^A}\frac{r(z_s)-r(z)}{r(z_s)} ~ .
\label{eq:lweight}
\ea
The weak lensing tomography provides us with a projected density
estimator for each redshift bin of the source galaxies, and hence possesses the significant
cross-correlations with the galaxy distribution observables.

\subsubsection{The redshift distribution of source galaxies}

We here assume the redshift distribution of sample galaxies is give by
the following analytic form \cite{Smail:1994,Amara:2007};
\begin{equation} 
    N(z) \propto z^{\alpha}
        \exp \left[ -\left(\frac{z}{z_0} \right)^{\beta} \right] ,
\label{eq:dNdz}
\end{equation} 
where $N(z)$ should be normalized as $\int N(z) dz = 1$, and we can
then rewrite the redshift distribution function of sample galaxies as,
\begin{equation}
    N(z) = \frac{n(z)}{\bng}  
    = \frac{\beta}{\Gamma[(\alpha+1)/\beta]} 
      \frac{z^{\alpha}}{z_0^{\alpha+1}}
        \exp \left[ -\left(\frac{z}{z_0} \right)^{\beta} \right] .
\label{eq:nz}
\end{equation}
We adopt
$\alpha=2.0$, $\beta=1.5$, and $z_0$ is determined from the relation with
mean redshift $z_{\rm m}$ defined as
\begin{equation}
    z_{\rm m} = \int z \frac{n(z)}{\bng} dz 
      = z_0 \frac{\Gamma[(\alpha+2)/\beta]}{\Gamma[(\alpha+1)/\beta]} .
\end{equation}
The normalization of the redshift
distribution function is fixed by the total number density of sample galaxies and defined as, 
\begin{equation}
  n^{\rm A} \equiv \int_0^\infty dz n(z) . 
\label{eq:nA}
\end{equation}

We next consider the tomographic surveys, which divide the sample
galaxies into some redshift bins, and analytically include the effect
of photometric redshift errors following the model of \cite{Ma:2006}
as
\begin{equation}
    p_i(z_{\rm ph} | z) = \frac{1}{\sqrt{2\pi}\sigma_z(z)}
        \exp \left[ -\frac{ \left(z-z_{\rm ph}\right)^2}
        {2\sigma_z^2(z)} \right] ,
\end{equation}
where $\sigma_z(z)$ denotes the redshift scatter systematics 
and we assume the following redshift dependence:
\begin{eqnarray}
    \sigma_z(z) = \sigma_z^{(i)}(1+z) , 
\end{eqnarray}
In this paper we assume only the redshift scatter systematics with
$\sigma_z^{(i)}=0.03$ for each redshift bin. We can then write the
redshift distribution of sample galaxies in the $i$-th redshift bin
with the effect of photometric redshift errors as
\begin{eqnarray}
    n_i(z)
    = \int_{z_{\rm ph}^{(i)}}^{z_{\rm ph}^{(i+1)}} dz_{\rm ph}
        N(z) p_i(z_{\rm ph} | z) 
    = \frac{1}{2} \left[ {\rm erf}(x_{i+1}) - {\rm erf}(x_i) \right] n(z) ,
\label{eq:numg}
\end{eqnarray}
with 
$x_i \equiv ( z_{\rm ph}^{(i)} -z )/\sqrt{2}\sigma_z(z)$ and the
redshift distribution of galaxy samples $n(z)$ defined in
Eq.~(\ref{eq:nz}). In the same manner with Eq.~(\ref{eq:nA}), the total
number density of sample galaxies in the $i$-th redshift bin can be
given as,
\begin{equation}
  n^{\rm A}_i \equiv \int_0^\infty dz n_i(z) . 
\label{eq:nAi}
\end{equation}
\\

In the following analysis, we assume a fixed number of photometric
redshift bins ($N^{\rm ph}=5$), and we determine the ranges
of redshift for each redshift bin keeping the same number of sample
galaxies for each bin. To see the dependence of the binning scheme on our
parameter estimation, we compare the parameter constraints with
different numbers of redshift bins in the discussion section Sec.~\ref{sec:discuss}.

\subsection{Three-dimensional galaxy power spectrum}
\label{sec:pkg}

In analyzing the three-dimensional redshift space galaxy power
spectrum, we restrict our discussions to the linear scale unless
stated otherwise.  The galaxy over-density $\delta_{\rm g}$ is distorted
along the line of sight due to the coherent infall bulk motion of
galaxies, and it has, in the plane parallel approximation (the
galaxies are far away from the observers so that the displacement due
to the peculiar velocities are parallel to each other), a simple form
in the Fourier space \cite{Kaiser:1987}, 
\ba
  \delta_{\rm g}(k,\mu)=b(z) \left[ 1+\beta(z,k) \mu^2 \right] \delta(k,z) ~,
\ea
where $b(z)$ is the scale-independent linear bias parameter and 
$\beta$ is the redshift space distortion parameter defined as
\ba
  \beta(z,k) = \frac{1}{b(z)} \frac{\partial \ln D(z,k)}{\partial \ln a} ~.
\ea
The magnitude of the wave number is $k=\sqrt{k_{\bot}+k_{\parallel}}$,
where $k_\bot$ and $k_\parallel$ are the wave numbers across and along
the line of sight, and $\mu$ represents the cosine of the angle
$\theta$ between the line of sight and the wave vector, $\mu_{}^2
\equiv \cos^2 \theta =k_{ \parallel}^2/(k_{\parallel}^2+k_{ \bot}^2)$.

In addition, the wrong assumption of the reference cosmological model also
produces the distortion in the measured power spectrum, known as
the geometrical distortion
\cite{Alcock:1979,Matsubara:1996,Ballinger:1996}. What we need in the measurements of
three-dimensional galaxy power spectrum is the positions in
the 3-dimensional space, though we measure the angular positions of
galaxies on the sky and the radial positions of galaxies in the redshift
space. We therefore need assume a reference cosmological model for
the mapping of the observed angular and redshift positions to
the one in the 3-dimensional space. We take the reference cosmology to be the
fiducial cosmology for simplicity in the following.

The comoving size of an object in the radial $r_{\parallel}$ and transverse
$r_{\bot}$ directions, which extends $\Delta \theta$ in the
angle and $\Delta z$ in the redshift, is estimated as
\ba
r_{\parallel} = \frac{\Delta z}{H(z)} ~, 
\hspace{5pt} {\rm and} \hspace{10pt}
r_{\bot} = D_{\rm A}(z) \Delta \theta ~,
\ea
where we need to assume some cosmological model for the estimation of $H(z)$
and the angular diameter distance $ D_{\rm A}(z)$.
Then the relations between the fiducial wave
numbers and the true wave numbers are estimated from the inverse of
$r_{\parallel}$ and $r_{\bot}$   
\ba
  k_{\parallel} = \frac{H(z)}{H^{({\rm fid})}(z)} k_{\parallel}^{({\rm fid})} ~, 
\hspace{5pt} {\rm and} \hspace{10pt}
  k_{\bot} = \frac{D_{\rm A}^{({\rm fid})}(z)}{D_{\rm A}(z)} k_{\bot}^{({\rm fid})} ~, 
\ea
where we distinguish the quantities in the fiducial cosmological model by the subscript ``fid''. 
Then the galaxy power spectrum estimated in redshift space is
modeled as \cite{Seo:2003}
\ba
\label{rsd}
  P_{\rm g}(k_{\parallel}^{({\rm fid})}, k_{\bot}^{({\rm fid})}, z)
  = \frac{D_{\rm A}^{({\rm fid})}(z)^2 H(z)}{D_{\rm A}(z)^2 H^{({\rm fid})}(z)} 
  b(z)^2 \left[1+ \beta(z,k) \mu_{}^2 \right]^2  P_{\rm m}(k,z) , 
\label{eq:pkg}
\ea
where the factor ${D_{\rm A}^{({\rm fid})}(z)^2 H(z)}/{D_{\rm A}(z)^2
  H^{({\rm fid})}(z)}$ represents the geometrical distortion
\cite{Alcock:1979}.
We can see from Eq.~(\ref{eq:pkg}) that using the galaxy power spectrum for
the cosmological parameter estimation would suffer from the degeneracies among the biasing
parameter $b(z)$, the redshift distortion parameter $\beta(z,k)$ and
the growth factor of the matter density fluctuations $D(z,k)$. 
As we shall discuss in the following
sections, these degeneracies can be lifted by
taking the cross correlations with other measurements. 
\\

We use the publicly available Boltzmann code CAMB \cite{Lewis:2000} to
get the transfer functions of the CMB observables and matter power
spectrum with massive neutrinos and calculate the corresponding all
auto- and cross-correlations.


\section{Fisher matrix formalism}
\label{sec:fisher}

For forecasting the cosmological parameter estimations, we calculate
the Fisher matrix (FM) which represents the curvature of the
likelihood function around its maximum in a given parameter space
defined as
\ba
  F_{\alpha\beta}\equiv \left\langle - \frac{\partial^2 \, {\ln L}}{\partial p_{\alpha} \partial p_{\beta}} \right\rangle
\ea
where $L$ is the likelihood function for the data set of our interest
given the theoretical model parameters ${\bm p}=\{p_1,\,p_2,...\,\}$
and the sub-scripts $\alpha$ and $\beta$ run over the model
parameters.
The FM formalism tells us how accurately the given observation can
measure the cosmological parameters around the fiducial model, and the
marginalized one-sigma error bound for a parameter $p_{\alpha}$ is
given by $\sigma(p_{\alpha})=\sqrt{({\bm F}^{-1})_{\alpha \alpha}}$ from the Cramer-Rao inequality.

As for our our fiducial cosmological model, we assume the flat
$\Lambda$CDM plus massive neutrino model in our analysis and we
include the following 13 cosmological parameters;
\begin{eqnarray}
{\bm p} \in
\{
100\Omega_{\rm b}h^2,  \;
\Omega_{\rm c}h^2,  \;
\Omega_{\Lambda},   \;
f_{\nu},           \;
\tau,             \;
\Yp,              \;
n_{\rm S},         \;
A_{\rm S},         \;
\alpha_{\rm S},    \;
w,                \;
N_{\rm eff},       \; 
b_{0},       \; 
b_{1}        \; 
\} ~ .
\label{eq:params}
\end{eqnarray}
$\Omega_{\rm b}$, $\Omega_{\rm c}$ and $\Omega_{\Lambda}$ are
respectively the
density parameters of the baryon, CDM and the cosmological constant
$\Lambda$. $h\equiv H_0/100$ represents the Hubble parameter.
$f_\nu$ is the fraction of the massive neutrinos in the matter
component, defined as $f_\nu \equiv \Omega_\nu/\Omega_{\rm m}$ and
$\Omega_\nu$ is the density parameter of the massive neutrino.
$\tau$ is the optical-depth at the epoch of reionization. 
$\Yp$ is the fractional primordial abundance of helium. 
$A_{\rm S}$, $n_{\rm S}$ and $\alpha_{\rm S}$ are the amplitude, the
spectra-index and the running parameter of the primordial power
spectrum and we choose the pivot-scale at $k_0=0.05 \; \Mpc$.
$w$ is the equation of state parameter of the dark energy. 
$N_{\rm eff}$ is the effective number of the radiative components. 
$b_0$ and $b_1$ are the model parameters for the galaxy biasing parameter. 
The fiducial values of our choice are
\begin{eqnarray}
{\bm p}_{\rm fid} &=& 
\{2.205,\,0.1199,\,0.6817,\,{f_{\nu}},\,0.089,\,0.2477,\,0.9603,\,2.196,\,0,\,-1,\,3.046,\,1.0,\,0.8 
\} ~.  \nn \\
\end{eqnarray}
Our analysis estimates the parameter constraints for the fiducial models with
the different values of the lightest neutrino mass $m_{\nu,{\rm
    min}}$. For instance, the fiducial values of $f_{\nu}\equiv
{\Omega_\nu}/{\Omega_{\rm m}}$ are, for the normal (inverted) mass
hierarchy scenarios, $0.0012(0.0016)$, $0.00073(0.0012)$, $0.00062(0.0010)$
respectively for $m_{\nu,{\rm min}}$ values of $0.05$ eV, $0.01$ eV and $10^{-4}$ eV
(this last example value $10^{-4}$ eV can be interpreted as the
illustration for the massless
minimal neutrino mass). 
Throughout this paper, we assume a spatially flat Universe and the
Hubble parameter is adjusted to keep our Universe flat when we vary
the other cosmological parameters. 
\\

We first outline the Fisher matrix for the transverse modes,
representing the correlations on the projected celestial surface, in
terms of the angular power spectrum $C_\ell$ fully including the auto-
and cross-correlations between different observables: the CMB
temperature anisotropies, $E$-mode polarization, CMB lensing potential
($T$, $E$, $\psi$), the galaxy distributions (${\rm g}_i$) and the
galaxy weak lensing shear ($\gamma_i$). 
We then briefly review the Fisher matrix for the three-dimensional galaxy
power spectrum $P_{\rm g}({\bm k})$.

\subsection{The Fisher matrix for the angular power spectrum}

The Fisher matrix for the angular power spectrum is given by \cite{jain1,Tegmark:1997}.
\ba 
  F_{\alpha\beta} = 
  \sum _{\ell_{\rm min}}^{\ell_{\rm max}} 
  \frac{f_{\rm sky} (2\ell+1)}{2} {\rm Tr} \left[ 
  \frac{\partial {\bf C}_{\ell}}{\partial p_\alpha}
  {\bf C}_\ell^{-1}
  \frac{\partial {\bf C}_{\ell}}{\partial p_\beta} 
  {\bf C}_\ell^{-1} \right] ~,
\label{eq:fisher_cls}
\ea
where $\fsky$ is the sky coverage of given survey and ${\bf C}_\ell$
represents the covariance matrix for the angular power spectra.

For the angular power spectrum, we consider the CMB and the photometric
redshift survey observables ($T, E,\psi,\{{\rm g}\},\{\gamma\}$).
We take into account the tomographic survey, and $\{{\rm g}\}=\{{\rm
  g}_1,..,{\rm g}_{N_{\rm g}^{\rm ph}} \}$ and
$\{\gamma\}=\{\gamma_1,..,\gamma_{N_\gamma^{\rm ph}}\}$ represent the
spectra from the tomographic redshift bins; $N^{\rm ph}_{\rm g}$ and
$N^{\rm ph}_\gamma$ are the number of the photometric redshift bins
for the galaxy clustering and the galaxy weak lensing observables,
respectively.

To take account of the different sky coverage for the CMB and the
photometric redshift surveys, we define the total Fisher matrix for
the angular power spectrum as \cite{Takeuchi:2012};
\begin{equation}
  F_{\alpha \beta}^{\rm 2D} = 
  \fsky^{\rm LSS} \sum_{\ell=2}^{\ell_{\rm max}^{\rm LSS}} {\cal F}_{\ell,\alpha \beta}^{\rm cross} 
  + (\fsky^{\rm CMB}-\fsky^{\rm LSS}) \sum_{\ell=2}^{\ell_{\rm max}^{\rm LSS}} {\cal F}_{\ell,\alpha \beta}^{\rm CMB}
  + \fsky^{\rm CMB} \sum_{\ell_{\rm max}^{\rm LSS}+1}^{\ell_{\rm max}^{\rm CMB}} {\cal F}_{\ell,\alpha \beta}^{\rm CMB} ~,
\end{equation}
with
\begin{eqnarray}
  {\cal F}_{\ell,\alpha \beta}^{\rm cross} &\equiv&
\frac{(2\ell+1)}{2} {\rm Tr} \left[ 
  \frac{\partial {\bf C}_{\ell}^{\rm cross}}{\partial p_\alpha}
  ({\bf C}_\ell^{\rm cross})^{-1}
  \frac{\partial {\bf C}_{\ell}^{\rm cross}}{\partial p_\beta} 
  ({\bf C}_\ell^{\rm cross})^{-1} \right] ~ , \\
  {\cal F}_{\ell,\alpha \beta}^{\rm CMB} &\equiv&
\frac{(2\ell+1)}{2} {\rm Tr} \left[ 
  \frac{\partial {\bf C}_{\ell}^{\rm CMB}}{\partial p_\alpha}
  ({\bf C}_\ell^{\rm CMB})^{-1}
  \frac{\partial {\bf C}_{\ell}^{\rm CMB}}{\partial p_\beta} 
  ({\bf C}_\ell^{\rm CMB})^{-1} \right] ~,
\end{eqnarray}
where
\ba
{\bf C}_\ell^{\rm cross} =
 \left( \begin{array}{llllll}
C_\ell^{TT}+N_\ell^{T} & C_\ell^{TE} & C_\ell^{T \psi} &C_\ell^{T \{\gamma\}}& C_\ell^{T \{{\rm g}\}} \\
C_\ell^{TE}& C_\ell^{EE}+N_\ell^{E} & 0 & 0 &0\\
C_\ell^{T \psi} & 0  & C_\ell^{\psi \psi}+ N_\ell^{\psi} & C_\ell^{\psi \{\gamma\}}&  C_\ell^{\psi \{{\rm g}\}}\\
C_\ell^{T\{\gamma\}} & 0 & C_\ell^{\psi \{\gamma\}} &  C_\ell^{\{\gamma\} \{\gamma\}}+ N_\ell^{\{\gamma\}} &   C_\ell^{\{\gamma\} \{{\rm g}\}} \\
C_\ell^{T \{{\rm g}\}} & 0 & C_\ell^{\psi \{{\rm g}\}} &  C_\ell^{\{\gamma\} \{{\rm g}\}} &  C_\ell^{\{{\rm g}\} \{{\rm g}\}}+ N_\ell^{\{{\rm g}\}} 
\end{array} \right) \; ~, \nn \\
\label{eq:cov2D_cross}
\ea 
and 
\ba
{\bf C}_\ell^{\rm CMB} =
 \left( \begin{array}{lll}
C_\ell^{TT}+N_\ell^{T} & C_\ell^{TE} & C_\ell^{T \psi} \\
C_\ell^{TE}& C_\ell^{EE}+N_\ell^{E} \\
C_\ell^{T \psi} & 0  & C_\ell^{\psi \psi}+ N_\ell^{\psi} 
\end{array} \right) \; ~ .
\label{eq:cov2D_CMB}
\ea 
Here the different super-scripts ``CMB'' and ``LSS'' represent the
variables for the CMB and the photometric redshift surveys,
respectively. The label ``cross'' represents the overlap region
between the CMB and the photometric redshift surveys and we assume
that the photometric redshift survey fully overlaps with the CMB
survey.
The covariance
matrix ${\bf C}_\ell$ and ${\bf C}_\ell$ include both signal $C_\ell^{XY}$ and noise

We assume that the cross-correlation between CMB $E$-mode
polarization and the other observables except $T$ should be small
because most of $E$-model polarization is generated through the
Thomson-scattering at the last scattering surface and not produced in the late time Universe
\footnote{During the epoch of reionization, some $E$-mode polarization
  can be produced and it can correlate with the CMB lensing potential or the
  objects at the high redshifts \cite{Lewis:2011,Dvorkin:2009a}.}.

For the CMB noise spectra, we simply consider the dominant
detector noise represented by the photon shot noise for a single
channel \cite{knox95,bond97}
\ba
  N_{\ell,\nu}^{T,E}=
  \left(\theta_{\rm FWHM} \Delta_{\nu}^{T,E} \right)^{2}
  \exp \left[( {\ell(\ell+1)\theta^2_{\rm FWHM}/8\ln2} \right] ,
\ea
where $\theta_{\rm FWHM}$ is the beam size or the spatial resolution of the beam and 
$\Delta_{\nu}^X$ represents the sensitivity of each channel to the
temperature $\Delta_{\nu}^{T}$ or polarization
$\Delta_{\nu}^P=\Delta_{\nu}^E=\Delta_{\nu}^B$, and these values are
summarized in Table~\ref{tb:survey_CMB}.
The corresponding noise power spectrum for a multi-channel
experiment is obtained by adding the contribution from all the
channels 
$N_{\ell}=\left[ \sum_\nu N_{\ell,\nu}^{-1} \right]^{-1}$.
The noise of the CMB lensing potential $N_\ell^{\psi}$ is estimated as
the statistical error of the lensing reconstruction based on the
optimal quadratic estimator \cite{Okamoto:2003}. 
\\

\begin{table}[t]
\begin{center}
\begin{tabular}{cccccccccc}
\hline \hline
 & 
 \makebox[50pt][c]{$f_{\rm sky}$} & 
 \makebox[50pt][c]{$\nu$} & 
 \makebox[50pt][c]{$\theta_{\rm FWHM}$} & 
 \makebox[60pt][c]{$\Delta_\nu^{\rm T}$} & 
 \makebox[60pt][c]{$\Delta_\nu^{\rm P}$}  \\
 Experiment &  & [GHz] & [arcmin] & [$\mu$K/pixel] & [$\mu$K/pixel]  \\
\hline
Planck & 0.65 
  & 100 & 9.5' & 6.8 & 10.9   \\
  && 143 & 7.1' & 6.0 & 11.4  \\
  && 217 & 5.0' & 13.1 & 26.7 \\
\hline \hline
\end{tabular}
\end{center}
\caption{
  The specifications for the CMB experiment.
  $\fsky$ denotes the fractional sky coverage, $\nu$ is the channel frequency,
  $\theta_{\rm FWHM}$ is the beam width and $\Delta_T$ ($\Delta_P$) represents the sensitivity of each channel to
  the temperature (polarization).
}
\label{tb:survey_CMB}
\end{table}
\begin{table}[t]
\begin{center}
\begin{tabular}{cccccc}
\hline \hline
 \makebox[25mm][c]{Survey  } &
 \makebox[25mm][c]{$\fsky$ } &
 \makebox[25mm][c]{$\bng$[arcmin$^{-2}$]  } &
 \makebox[25mm][c]{$\zm$   } &
 \makebox[25mm][c]{$\bar{\sigma}_\gamma$   } \\
\hline
  DES  & 0.1 & 10 & 0.8 & 0.3  \\
  HSC  & 0.05 & 30 & 1.0 & 0.3  \\
  LSST & 0.5  & 50 & 1.2 & 0.3 \\
\hline \hline
\end{tabular}
\end{center}
\caption{
  The specifications for the photometric redshift surveys. 
  $\fsky$ denotes the fractional sky coverage, 
  $\bng$ is the surface number density of sample galaxies, 
  $\zm$ is the mean redshift of the survey and 
  $\bar{\sigma}_\gamma$ denotes the error of shear measurements. 
}
\label{tb:survey_phz}
\end{table}

The observed shear power spectrum is contaminated by
the shear variance per galaxy 
$\bar{\sigma}_\gamma^2 $  
due to the
uncertainties in the intrinsic shape of the source galaxies
\ba
N^{\gamma_i}_\ell
=\frac{\bar{\sigma}_\gamma^2 }{\bar{n}_{{\rm g},i}} \delta_{ij} \; .
\ea
Analogously, the noise for the galaxy angular power spectrum is
\ba
N^{{\rm g}_i}_\ell =\frac{ \delta_{ij} }{\bar{n}_{{\rm g},i}} \; ,
\ea
where $\bar{n}_{{\rm g},i}$ is the number density of sample galaxies per
steradian in the $i$-th tomographic redshift bin and we assume the
noise is uncorrelated among different tomographic bins. 
Throughout this paper, we divide the photometric redshift bins in such
a way that the number density of sample galaxies of each redshift bin
should be same, $\bar{n}_{{\rm g},i}=\bar{n}_{\rm g}/N^{\rm ph}$.
Table~\ref{tb:survey_phz} shows the specifications for the photometric
redshift surveys used in our analysis.  \\

\begin{table}[t]
\begin{center}
\begin{tabular}{ccccccccccc}
\hline \hline
 \makebox[10mm][c]{$i$} &
 \makebox[20mm][c]{$\zc$}  & 
 \makebox[25mm][c]{$\bng^{\rm 3D}$} &
 \makebox[25mm][c]{$\Vs/\fsky$} &
 \makebox[25mm][c]{$\knl$} 
\\
 &
 &
 [$10^{-3}h^3 \Mpc^{-3}$] &
 [$h^{-3}\Gpc^{-3}$] &
 [$h^{-1}\Mpc$] \\
\hline
 1 & 0.26 & 33.0 & 10.6 &  0.12 \\
 2 & 0.61 & 34.7 & 11.8 &  0.15 \\
 3 & 0.77 & 35.2 & 13.2 &  0.16 \\
 4 & 0.92 & 35.5 & 14.7 &  0.18 \\
 5 & 1.06 & 35.6 & 16.7 &  0.20 \\
 6 & 1.20 & 35.6 & 18.4 &  0.21 \\
 7 & 1.35 & 35.3 & 22.7 &  0.23 \\
 8 & 1.53 & 34.7 & 30.3 &  0.25 \\
 9 & 1.77 & 33.4 & 44.9 &  0.29 \\
10 & 2.45 & 27.6 &  182 &  0.40 \\
\hline \hline
\end{tabular}
\end{center}
\caption{ 
  The survey parameters for the Euclid spectroscopic galaxy
  redshift survey in our analysis.  $i$ denotes the values in the
  $i$-th redshift slice, $\zc$ is the central redshift in each
  redshift bin, $\Vs$ and $\bng^{\rm 3D}$ are the comoving survey
  volume and number density of sampled galaxies, and 
  $k_{\rm nl}=\pi/(2R)$ is the wave-number corresponding to the the
  non-linear scales with the criteria of $\sigma_R(z)=0.5$. We here
  adopt the fractional sky coverage $\fsky=0.2$. }
\label{tb:survey_pkg}
\end{table}

\subsection{The Fisher matrix for the three-dimensional galaxy power spectrum}

Each redshift slice of the spectroscopic redshift survey provides
an independent information and the correlations between different
redshift slices should be negligible. 
The Fisher matrix for the
galaxy power spectrum $P_{\rm g}({\bm k})$ from the spectroscopic
redshift survey then reads \cite{Tegmark:1997,nick2,feld,teg,seo1};
\ba
  F_{\alpha \beta}^{{\rm 3D}}
  = \sum_i^{N^{\rm sp}}
  \frac{V_{s,i}}{8 \pi^2}\int^{1}_{-1} d \mu \int ^{k_{{\rm max}}}_{k_{{\rm min}}} k^2 dk 
  \frac{\partial P_{\rm g}(k,\mu,z_{c,i})}{\partial p_{\alpha}}
  \frac{\partial P_{\rm g}(k,\mu,z_{c,i})}{\partial p_{\beta}} 
  \left[\frac{P_{\rm g}(k,\mu,z_{c,i})}{P_{\rm g}(k,\mu,z_{c,i})+1/\bar{n}_{{\rm g},i}^{\rm 3D}}\right]^2 ~, \nn \\
\label{eq:fisher_pkg}
\ea
where $N^{\rm ph}$ is the number of the redshift slices, $z_c$ is
the central redshift of the redshift bin, $\bar{n}_g^{\rm 3D}$ and $V_s$ are
respectively the mean comoving galaxy number density and the comoving
volume of the galaxy survey, and the sub-script $i$ denotes the value
in the $i$-th redshift slice. We summarize these values in each
redshift slice in Table~\ref{tb:survey_pkg}, and these estimations are
carried out as follows.

The mean comoving number density of the sample galaxies is determined as
\begin{equation}
  \bar{n}_{{\rm g},i}^{\rm 3D} = \int_{\Mmin}^{\infty} dM \nh(M,z_{c,i}) \langle N \rangle_{M} ,
\label{eq:ng_sp}
\end{equation}
where the functions $\nh$ is the halo
mass function for a given mass $M$ and a redshift $z$. 
We use the mass function $\nh(M,z)$ given by Warren
et.\,al. \cite{Warren:2006}. 
$\langle N\rangle_M$ is the halo-occupation distribution describing how many galaxies
the halo with th mass $M$ hosts, and we here simply assume $\langle N
\rangle_M = 1$. The parameter $\Mmin$ is the minimum halo mass hosting
the observed galaxies, which
corresponds to the observational limits of each survey. We here simply
set $\Mmin=10^{11.0}\Msun/h$ for all redshifts.

The comoving survey volume is approximately defined as
\begin{equation}
  V_{{\rm s},i} \simeq 4\pi \fsky r(z_{c,i})^2 \frac{dr}{dz} \Delta z_i
\end{equation}
where $\Delta z_i$ denotes the width of $i$-th redshift slice.

For the range of the $k$-integration, we set the maximum wave-number
$k_{\rm max}$ to avoid using the nonlinear $k$ modes. We used the criteria for the linear scale $k< k_{\rm max}$ by
demanding the amplitude of the
smoothed density field $\sigma_R(z)<0.5$, so that $k_{\rm max}=\pi/(2R)$ for $\sigma_{R}(z)=0.5$. The variance of
the smoothed density field here is
given by
\begin{equation}
  \sigma_R^2(z) = \frac{1}{2\pi^2} \int k^2 P_{\rm m}(k) W(kR)^2 \frac{D(z)^2}{D(0)^2} dk , 
\end{equation}
The function $W(kR)$ is the top-hat window function, $R$ is the
smoothing scale, and $D_0$ is the growth factor at the present
time. 
For the minimum wave-number, we simply choose
$k_{\min}=10^{-4}$ $h$/Mpc for all the redshift slices.  
\\

In combining the Fisher matrix of $P_{\rm g}({\bm k})$ including both
transverse and longitudinal mode information and that of
$C_l$ including only the transverse mode information, to avoid the complications in the full covariance matrix, we
make a simplified assumption that their transverse modes are fully correlated and take
account of the common transverse modes only once
\cite{gata,cai,taka2}. 
Accordingly, in our combining the spectroscopic galaxy redshift survey
with the photometric galaxy survey for an overlapping sky area, we
remove the section around $\mu=0$ 
in integrating the Fisher matrix over $\mu$ not to over-count the
transverse modes in estimating the parameter uncertainties 
\footnote{More rigorous treatment to cross-correlate the
  three-dimensional and two-dimensional data will be presented in our
  separate forthcoming paper.}. 
This hence would give us a conservative estimation for the total
information by combining these measurements and the actual cosmological
parameter uncertainties would be smaller due to the smaller cross
correlations among the common transverse modes in $P_{\rm g}({\bm k})$ and
$C_\ell$. \\

\begin{table}[t]
\begin{center}
\begin{tabular}{clccccc}
\hline \hline
  \makebox[25mm]{Survey}
 & \makebox[30mm]{Observables}
 & \makebox[10mm]{$\fsky$}
 & $\ell_{\rm min}$ or $k_{\rm min}$
 & $\ell_{\rm max}$ or $k_{\rm max}$
 & redshift bin \\
\hline
  Planck & CMB       ($T$, $E$, $\psi$) & 0.65 & 2 & 3000 & --- \\
  LSST   & photo-$z$ (${\rm g}$, $\gamma$) & 0.5 & 2 & 500 & 5 \\
  Euclid & spec-$z$  (${\rm g^{\rm 3D}}$) & 0.2 & $10^{-4}$ $h$/Mpc & $k_{\rm nl}$ & 10 \\  
\hline \hline
\end{tabular}
\end{center}
\caption{The reference survey model in our analysis. }
\label{tb:fiducial}
\end{table}

We put the constraints on the cosmological parameters by combining the
upcoming data from three kinds of the survey projects. The first is the CMB
experiment by the Planck satellite which provides 
the temperature anisotropies ($T$), the $E$-mode polarization ($E$) and the weak
lensing of the CMB ($\psi$; the lensing potential). The second is the photometric
redshift survey by the LSST which provides the information of
the galaxy clustering (${\rm g}$) and the galaxy weak lensing
($\gamma$; the lensing shear). The third is the spectroscopic redshift survey by the
Euclid which provides the 3-dimensional galaxy power spectrum
(${\rm g}^{\rm 3D}$). We use these projects as our fiducial survey models
and the survey parameters are summarized in Table~\ref{tb:fiducial}.


\section{Results}
\label{sec:result}

\begin{figure}[t]
\begin{center}
\includegraphics[clip,keepaspectratio=true,width=1.0
  \textwidth]{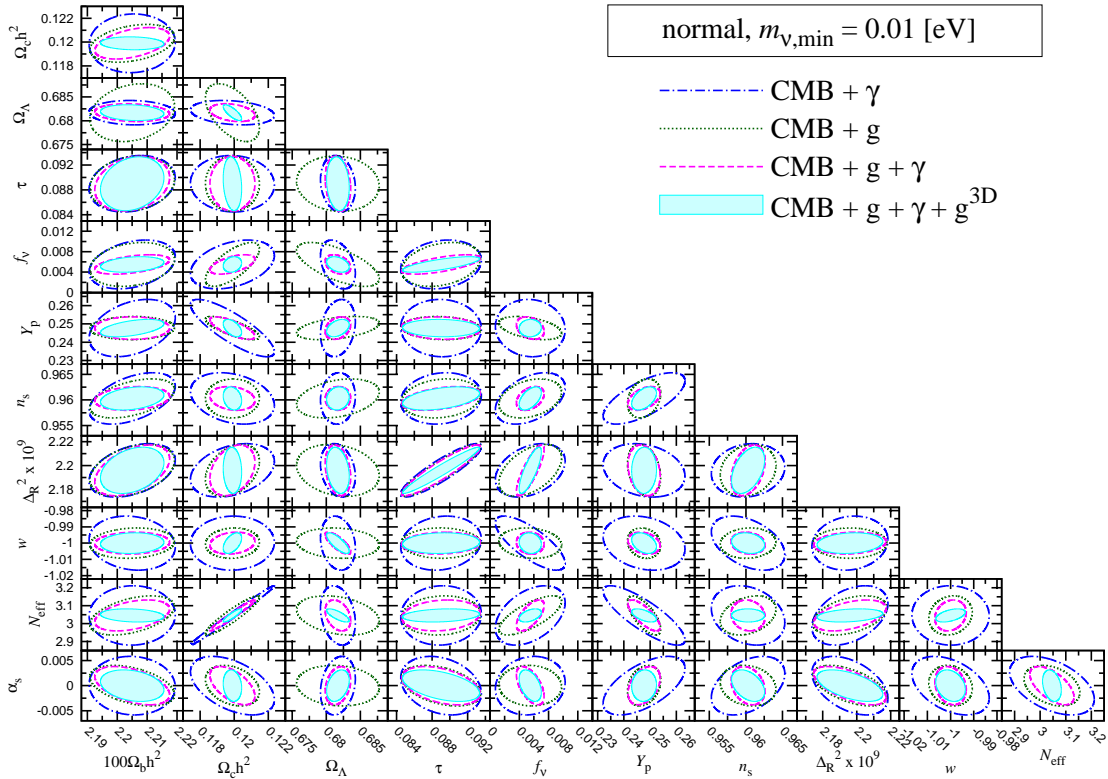}
\end{center}
\caption{
  The projected 1-$\sigma$ (68\%) CL areas on each
  parameter plane for the normal
  mass hierarchy scenario with the lightest neutrino 
  mass $m_{\nu,{\rm min}}=0.01$ eV. 
  We compare the four combinations of the observables from the Planck ($T,\,E,\,\psi$), LSST (${\rm g},\,\gamma$)
  and Euclid (${\rm g}^{\rm 3D}$); 
  1) (blue) CMB and galaxy weak lensing ($T,\,E,\,\psi,\,{\gamma}$), 
  2) (green) CMB and galaxy clustering ($T,\,E,\,\psi,\,{\rm g}$), 
  3) (magenta) CMB, galaxy clustering and weak lensing ($T,\,E,\,\psi,\,{\rm g},\,\gamma$), and 
  4) (cyan) CMB, galaxy clustering, weak lensing and 3-dimensional galaxy power spectrum ($T,\,E,\,\psi,\,{\rm
    g},\,\gamma,\,{\rm g}^{3D}$). 
}
\label{fig:cont_n}
\end{figure}

\begin{figure}[t]
\begin{center}
\includegraphics[clip,keepaspectratio=true,width=1.0
  \textwidth]{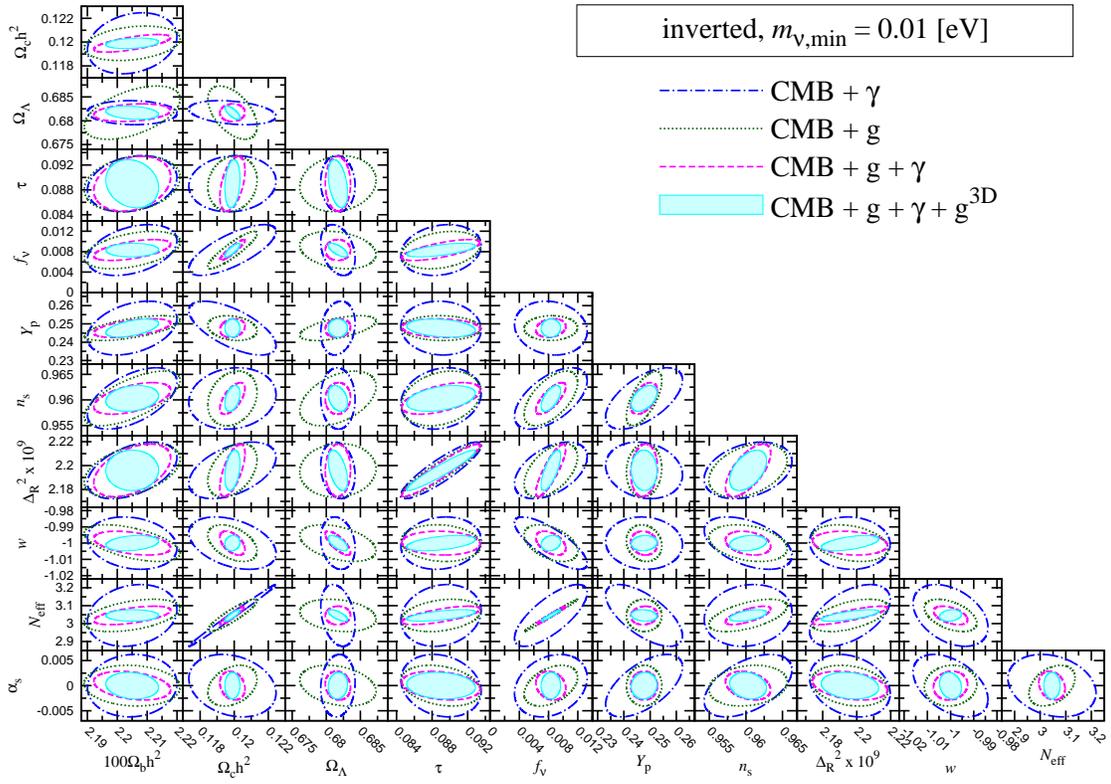}
\end{center}
\caption{
  Same as Figure~\ref{fig:cont_n}, but for the 
  inverted mass hierarchy scenario with the same value of the lightest neutrino mass. 
}
\label{fig:cont_i}
\end{figure}

\begin{figure}[t]
\begin{center}
\begin{minipage}[t]{0.75\textwidth}
\includegraphics[clip,keepaspectratio=true,width=0.95
  \textwidth]{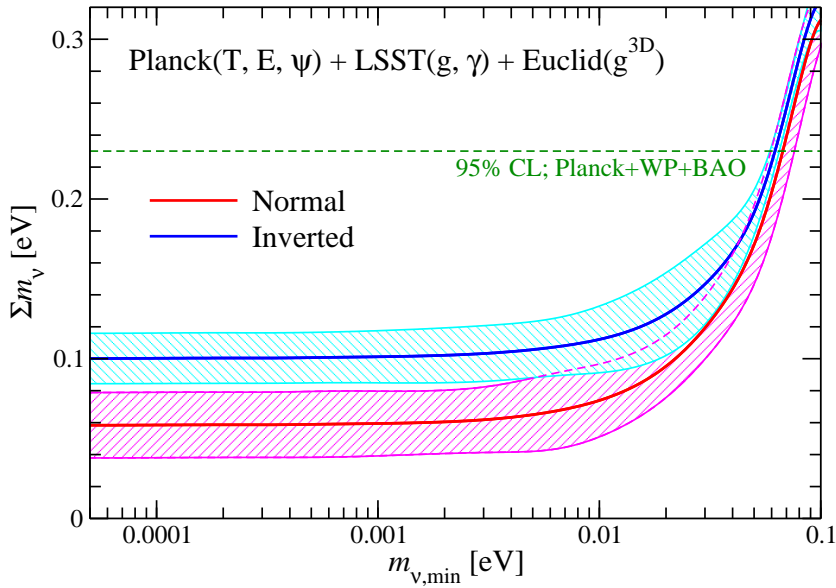}
\end{minipage}
\end{center}
\caption{The projected 1-$\sigma$ error for the total neutrino mass $\sum m_\nu$
  as a function of the lightest neutrino mass
  $m_{\nu,{\rm min}}$ for the normal (red) and inverted (blue) mass
  hierarchy scenarios. The
  thick solid curves representing the fiducial $\sum m_\nu$ for a
  given $m_{\nu,{\rm min}}$ are obtained using the mass difference values from
  the neutrino oscillation data (given by Eq.~(\ref{ndata})). The
  filled curves represent the 1-$\sigma$ CL regions estimated by combining 
  all the observables ($T$, $E$, $\psi$, ${\rm g}$, $\gamma$ and ${\rm g}^{\rm 3D}$), 
  and we assume Planck for the CMB, LSST for the
  galaxy clustering and the galaxy weak lensing observables, and Euclid for the 3-dimensional galaxy power spectrum. 
}
\label{fig:hierarchy}
\end{figure}

We here present the results of our forecasts for the cosmological
parameters illustrating the breaking of the parameter degeneracies by
combining the various observables.  Figures~\ref{fig:cont_n} and
\ref{fig:cont_i} show the projected 1-$\sigma$ confidence limit (CL)
contours on each parameter plane for, respectively, the normal and
inverted mass hierarchy scenarios with the lightest
neutrino mass $m_{\nu,{\rm min}}=0.01$ eV. 
These panels compare the constraints for four cases ; 
1) ``CMB + ${\gamma}$'' combines the CMB and galaxy weak lensing observables.  
2) ``CMB + ${\rm g}$'' combines the CMB and galaxy clustering observables. 
3) ``CMB + ${\rm g}$ + $\gamma$'' combines the CMB, galaxy weak lensing and galaxy
clustering observables. 
4) ``CMB + ${\rm g}$ + $\gamma$ + ${\rm g}^{\rm 3D}$'' combines the CMB, galaxy clustering, galaxy
weak lensing observables plus the 3-dimensional galaxy power spectrum.
We adopt the Planck for the CMB, the LSST for the galaxy weak lensing
and the photometric galaxy redshift survey, and the Euclid for the
spectroscopic galaxy redshift survey. We take into account all the
auto- and cross-correlations among these observables.
Table~\ref{tb:error} summarizes the projected 1-$\sigma$ error of each
parameter for the normal and inverted mass hierarchy scenarios for a
few representative lightest neutrino masses 
$m_{\nu,{\rm min}}=10^{-4}$ and 0.01 and 0.05 eV
\footnote{$m_{\nu,{\rm min}}=10^{-4}$ eV essentially represents the
  massless lightest neutrino case as demonstrated in
  Figure~\ref{fig:hierarchy}.
}.


\begin{table}[h]
\begin{center}
{\bf Normal hierarchy: $m_{\nu,{\rm min}}=m_1 < m_2 \ll m_3$}
{
\tiny
\begin{tabular}{lccccccccccccc||c}
\hline \hline
\makebox[4mm][l]{Observables} &
\makebox[4mm][c]{$100\Omega_{\rm b}h^2$} &
\makebox[4mm][c]{$\Omega_{\rm c}h^2$} &
\makebox[4mm][c]{$\Omega_{\Lambda}$} &
\makebox[4mm][c]{$\tau$} &
\makebox[4mm][c]{$f_\nu$} &
\makebox[4mm][c]{$Y_{p}$} &
\makebox[4mm][c]{$n_{\rm S}$} &
\makebox[4mm][c]{$A_{\rm S} \times 10^{9}$} &
\makebox[4mm][c]{$w$} &
\makebox[4mm][c]{$N_{\rm eff}$} &
\makebox[4mm][c]{$\alpha$} &
\makebox[4mm][c]{$b_0$} &
\makebox[4mm][c]{$b_1$} &
\makebox[4mm][c]{$\sum m_{\nu}$} \\
\hline
\multicolumn{3}{l}{{\scriptsize ${m_{\nu,{\rm min}}=0.05}$ eV}}\\
\cl
 &     0.023 &     0.00410 &     0.0893 &     0.0045 &     0.0123 &     0.0170 &     0.0107 &     0.025 &     0.3436 &     0.265 &     0.0080 &     ---    &      ---   &     0.202 \\ 
\w
 &     1.247 &     0.04021 &     0.0042 &     ---    &     0.0282 &     ---    &     0.0770 &     0.617 &     0.0410 &     0.766 &     0.0293 &     ---    &      ---   &     0.707 \\ 
\g
 &     0.024 &     0.00581 &     0.0129 &     ---    &     0.0219 &     ---    &     0.0334 &     0.477 &     0.0123 &     0.626 &     0.0296 &     0.0757 &     0.0338 &     0.348 \\ 
\hdashline
\clw
 &     0.015 &     0.00278 &     0.0024 &     0.0045 &     0.0050 &     0.0155 &     0.0060 &     0.023 &     0.0181 &     0.187 &     0.0061 &     ---    &      ---   &     0.091 \\ 
\clg
 &     0.016 &     0.00148 &     0.0054 &     0.0044 &     0.0035 &     0.0071 &     0.0052 &     0.021 &     0.0094 &     0.104 &     0.0040 &     0.0153 &     0.0086 &     0.061 \\ 
\clgw
 &     0.013 &     0.00116 &     0.0017 &     0.0043 &     0.0019 &     0.0068 &     0.0031 &     0.021 &     0.0068 &     0.078 &     0.0031 &     0.0032 &     0.0035 &     0.035 \\ 
\clgwsm \\
\clgws
 &     0.013 &     0.00052 &     0.0014 &     0.0043 &     0.0014 &     0.0064 &     0.0030 &     0.019 &     0.0056 &     0.035 &     0.0026 &     0.0029 &     0.0031 &     0.024 \\ 
\hline \hline
\multicolumn{3}{l}{{\scriptsize ${m_{\nu,{\rm min}}=0.01}$ eV}}\\
\cl
 &     0.023 &     0.00408 &     0.0907 &     0.0045 &     0.0136 &     0.0170 &     0.0107 &     0.025 &     0.3570 &     0.265 &     0.0079 &     ---    &     ---    &     0.199 \\ 
\w
 &     1.397 &     0.04402 &     0.0054 &     ---    &     0.0188 &     ---    &     0.0713 &     0.479 &     0.0427 &     0.722 &     0.0268 &     ---    &     ---    &     0.417 \\ 
\g
 &     0.168 &     0.00778 &     0.0135 &     ---    &     0.0146 &     ---    &     0.0226 &     0.270 &     0.0135 &     0.449 &     0.0165 &     0.0664 &     0.0204 &     0.229 \\ 
\hdashline
\clw
 &     0.014 &     0.00246 &     0.0025 &     0.0045 &     0.0047 &     0.0156 &     0.0050 &     0.022 &     0.0166 &     0.165 &     0.0058 &     ---    &     ---    &     0.073 \\ 
\clg
 &     0.014 &     0.00158 &     0.0060 &     0.0044 &     0.0042 &     0.0063 &     0.0038 &     0.021 &     0.0093 &     0.109 &     0.0040 &     0.0135 &     0.0112 &     0.064 \\ 
\clgw
 &     0.012 &     0.00132 &     0.0018 &     0.0043 &     0.0018 &     0.0061 &     0.0023 &     0.021 &     0.0067 &     0.086 &     0.0038 &     0.0032 &     0.0039 &     0.030 \\ 
\clgwsm \\
\clgws
 &     0.010 &     0.00054 &     0.0016 &     0.0043 &     0.0015 &     0.0047 &     0.0022 &     0.019 &     0.0063 &     0.036 &     0.0031 &     0.0026 &     0.0031 &     0.022 \\ 
\hline \hline
\multicolumn{3}{l}{{\scriptsize ${m_{\nu,{\rm min}}=10^{-4}}$ eV}}\\
\cl
 &     0.023 &     0.00407 &     0.0896 &     0.0046 &     0.0126 &     0.0169 &     0.0106 &     0.025 &     0.3474 &     0.264 &     0.0079 &     ---    &     ---    &     0.181 \\ 
\w
 &     0.844 &     0.02851 &     0.0048 &     ---    &     0.0140 &     ---    &     0.0588 &     0.401 &     0.0430 &     0.692 &     0.0253 &     ---    &     ---    &     0.277 \\ 
\g
 &     0.172 &     0.00434 &     0.0181 &     ---    &     0.0098 &     ---    &     0.0333 &     0.260 &     0.0132 &     0.488 &     0.0126 &     0.1061 &     0.0322 &     0.149 \\ 
\hdashline
\clw
 &     0.015 &     0.00275 &     0.0031 &     0.0045 &     0.0049 &     0.0151 &     0.0058 &     0.023 &     0.0220 &     0.188 &     0.0058 &     ---    &     ---    &     0.074 \\ 
\clg
 &     0.016 &     0.00107 &     0.0051 &     0.0044 &     0.0029 &     0.0059 &     0.0054 &     0.021 &     0.0085 &     0.086 &     0.0041 &     0.0153 &     0.0066 &     0.042 \\ 
\clgw
 &     0.014 &     0.00090 &     0.0016 &     0.0043 &     0.0017 &     0.0052 &     0.0031 &     0.021 &     0.0065 &     0.059 &     0.0029 &     0.0031 &     0.0032 &     0.026 \\ 
\clgwsm \\
\clgws
 &     0.010 &     0.00048 &     0.0013 &     0.0042 &     0.0014 &     0.0051 &     0.0028 &     0.017 &     0.0060 &     0.029 &     0.0027 &     0.0030 &     0.0029 &     0.020 \\ 
\hline \hline
\end{tabular}
}

\vspace{10mm}

{\bf Inverted hierarchy: $m_{\nu,{\rm min}}=m_3 \ll m_1 < m_2$}
{
\tiny
\begin{tabular}{lccccccccccccc||c}
\hline \hline
\makebox[4mm][l]{Observables} &
\makebox[4mm][c]{$100\Omega_{\rm b}h^2$} &
\makebox[4mm][c]{$\Omega_{\rm c}h^2$} &
\makebox[4mm][c]{$\Omega_{\Lambda}$} &
\makebox[4mm][c]{$\tau$} &
\makebox[4mm][c]{$f_\nu$} &
\makebox[4mm][c]{$Y_{p}$} &
\makebox[4mm][c]{$n_{\rm S}$} &
\makebox[4mm][c]{$A_{\rm S} \times 10^{9}$} &
\makebox[4mm][c]{$w$} &
\makebox[4mm][c]{$N_{\rm eff}$} &
\makebox[4mm][c]{$\alpha$} &
\makebox[4mm][c]{$b_0$} &
\makebox[4mm][c]{$b_1$} &
\makebox[4mm][c]{$\sum m_{\nu}$} \\
\hline
\multicolumn{3}{l}{{\scriptsize ${m_{\nu,{\rm min}}=0.05}$ eV}}\\
\cl
 &     0.023 &     0.00413 &     0.0896 &     0.0045 &     0.01188 &     0.0170 &     0.0107 &     0.025 &     0.3416 &     0.267 &     0.0079 &     ---    &      ---   &     0.200 \\ 
\w
 &     1.193 &     0.03283 &     0.0049 &     ---    &     0.02943 &     ---    &     0.0492 &     0.347 &     0.0433 &     0.643 &     0.0269 &     ---    &     ---    &     0.715 \\ 
\g
 &     0.025 &     0.00484 &     0.0164 &     ---    &     0.00116 &     ---    &     0.0201 &     0.236 &     0.0566 &     0.378 &     0.0129 &     0.0420 &     0.0434 &     0.042 \\ 
\hdashline
\clw
 &     0.015 &     0.00346 &     0.0025 &     0.0044 &     0.00383 &     0.0131 &     0.0067 &     0.023 &     0.0168 &     0.213 &     0.0063 &     ---    &      ---   &     0.080 \\ 
\clg
 &     0.016 &     0.00174 &     0.0110 &     0.0036 &     0.00064 &     0.0077 &     0.0052 &     0.017 &     0.0390 &     0.124 &     0.0039 &     0.0240 &     0.0091 &     0.020 \\ 
\clgw
 &     0.012 &     0.00147 &     0.0023 &     0.0033 &     0.00064 &     0.0070 &     0.0023 &     0.013 &     0.0091 &     0.089 &     0.0032 &     0.0046 &     0.0047 &     0.018 \\ 
\clgwsm \\
\clgws
 &     0.012 &     0.00066 &     0.0019 &     0.0028 &     0.00061 &     0.0058 &     0.0022 &     0.012 &     0.0079 &     0.042 &     0.0029 &     0.0039 &     0.0033 &     0.014 \\ 
\hline \hline
\multicolumn{3}{l}{{\scriptsize ${m_{\nu,{\rm min}}=0.01}$ eV}}\\
\cl
 &     0.023 &     0.00407 &     0.0921 &     0.0046 &     0.0126 &     0.0170 &     0.0107 &     0.025 &     0.3574 &     0.263 &     0.0079 &     ---    &     ---    &     0.193 \\ 
\w
 &     1.323 &     0.03858 &     0.0049 &     ---    &     0.0230 &     ---    &     0.0810 &     0.543 &     0.0388 &     0.568 &     0.0243 &     ---    &     ---    &     0.529 \\ 
\g
 &     0.081 &     0.00494 &     0.0131 &     ---    &     0.0152 &     ---    &     0.0246 &     0.421 &     0.0395 &     0.364 &     0.0216 &     0.0454 &     0.0406 &     0.236 \\ 
\hdashline
\clw
 &     0.015 &     0.00255 &     0.0025 &     0.0045 &     0.0049 &     0.0146 &     0.0060 &     0.023 &     0.0161 &     0.172 &     0.0062 &     ---    &     ---    &     0.081 \\ 
\clg
 &     0.016 &     0.00144 &     0.0055 &     0.0044 &     0.0036 &     0.0067 &     0.0053 &     0.022 &     0.0112 &     0.090 &     0.0040 &     0.0142 &     0.0071 &     0.057 \\ 
\clgw
 &     0.012 &     0.00073 &     0.0018 &     0.0044 &     0.0020 &     0.0050 &     0.0030 &     0.022 &     0.0073 &     0.051 &     0.0028 &     0.0031 &     0.0032 &     0.032 \\ 
\clgwsm \\
\clgws
 &     0.010 &     0.00044 &     0.0014 &     0.0038 &     0.0013 &     0.0049 &     0.0025 &     0.016 &     0.0047 &     0.032 &     0.0026 &     0.0029 &     0.0029 &     0.020 \\ 
\hline \hline
\multicolumn{3}{l}{{\scriptsize ${m_{\nu,{\rm min}}=10^{-4}}$ eV}}\\
\cl
 &     0.023 &     0.00409 &     0.0919 &     0.0045 &     0.0128 &     0.0170 &     0.0107 &     0.025 &     0.3575 &     0.265 &     0.0079 &     ---    &     ---    &     0.194 \\ 
\w
 &     1.220 &     0.04075 &     0.0058 &     ---    &     0.0138 &     ---    &     0.0673 &     0.416 &     0.0420 &     0.670 &     0.0298 &     ---    &     ---    &     0.367 \\ 
\g
 &     0.267 &     0.01182 &     0.0157 &     ---    &     0.0064 &     ---    &     0.0406 &     0.159 &     0.0157 &     0.036 &     0.0098 &     0.0854 &     0.0435 &     0.143 \\ 
\hdashline
\clw
 &     0.013 &     0.00233 &     0.0026 &     0.0045 &     0.0030 &     0.0151 &     0.0045 &     0.021 &     0.0145 &     0.152 &     0.0059 &     ---    &     ---    &     0.051 \\ 
\clg
 &     0.015 &     0.00071 &     0.0050 &     0.0043 &     0.0020 &     0.0077 &     0.0045 &     0.019 &     0.0093 &     0.016 &     0.0036 &     0.0145 &     0.0104 &     0.031 \\ 
\clgw
 &     0.013 &     0.00037 &     0.0017 &     0.0037 &     0.0011 &     0.0059 &     0.0024 &     0.016 &     0.0066 &     0.014 &     0.0033 &     0.0035 &     0.0043 &     0.017 \\ 
\clgwsm \\
\clgws
 &     0.012 &     0.00035 &     0.0011 &     0.0036 &     0.0010 &     0.0058 &     0.0023 &     0.016 &     0.0053 &     0.013 &     0.0033 &     0.0030 &     0.0036 &     0.015 \\ 
\hline \hline
\end{tabular}
}
\end{center}
\caption{
  The projected 1-$\sigma$ (68\%) CL error of each parameter
  for the the normal (Top) and inverted (Bottom) mass hierarchy
  scenarios with the lightest neutrino
  masses of $m_{\nu,{\rm min}}=0.05$, $0.01$ and $10^{-4}$ eV. The first three
  rows respectively list the constraints from only the
  CMB($T,\,E,\,\psi$), the galaxy weak lensing ($\gamma$) and the
  galaxy clustering (${\rm g}$). The next four rows represent the
  results with different combinations of the observables. The last raw
  combines all the CMB, the galaxy weak lensing, the galaxy clustering
  observables and the 3-dimensional galaxy power spectrum (CMB + ${\rm g}$ +
  $\gamma$ + ${\rm g}^{\rm 3D}$(spec-$z$)).}
\label{tb:error}
\end{table}

A big advantage of adding the redshift survey to the CMB observables
is the breaking of the angular diameter distance degeneracies
\cite{eht98}. 
For the photometric redshift surveys, for instance, the galaxy weak
lensing information is helpful in breaking the $\nmsum$-$w$ degeneracy
\cite{hanne}. (This $\nmsum$-$w$ degeneracy arises indirectly from the
degeneracy between $w$-$\Omega_{\Lambda}$ and that between
$\nmsum$-$\Omega_{\Lambda}$. $w$ and $\Omega_{\Lambda}$ are
degenerate in affecting the late time Hubble expansion rate, while $\nmsum$ and $\Omega_{\Lambda}$ (or equivalently
$\Omega_{\rm m}$) are degenerate affecting the matter power spectrum
suppression scale.)  
The lensing weight $W_{\ell,i}(r)$ (Eq.~(\ref{eq:lweight})) is
sensitive to the dark energy parameters, $\Omega_{\Lambda}$ and $w$,
because the distance-redshift relation $r=\int dz/H(z)$ is affected
by the dark energy properties. Hence adding the galaxy weak lensing
information helps to constrain the dark energy parameters and it
consequently improves the neutrino mass constraints (even though the
light neutrinos of our interest would not directly affect the the
late time Hubble expansion rate). In particular, the tomographic
survey gives an additional handle on the distance-redshift relation
due to the different redshift dependence of $W_{\ell,i}(r)$ for each
redshift bin, which further helps in breaking the
$w$-$\Omega_{\Lambda}$ degeneracy.

While the neutrino masses are already tightly constrained from the
angular power spectra including the CMB and photometric redshift
survey data with all the cross correlations taken into account, we
found adding the three-dimensional galaxy power spectrum $P_{\rm
  g}({\bm k})$ still helps in further improving the neutrino mass
determination by as much as 20-40 \% depending on the neutrino mass structure as shown in
Table~\ref{tb:error}. This improvement stems from the reduction of the galaxy bias uncertainty by cross correlating the
different observables \cite{pen2,takada,Takeuchi:2012}.  From the
transverse mode information in the 2-dimensional angular correlations,
the galaxy bias can be inferred from the ratios of the galaxy power
spectrum (dependent on the bias squared), the lensing shear power
(independent of the bias) and their cross correlation (linearly
dependent on the bias), because they have the different dependence on
the bias. Adding the 3-dimensional galaxy power spectrum $P_{\rm
  g}({\bm k})$ with the redshift space distortion to the 2-dimensional
angular power spectra $C_l$ further reduces, in addition to adding
more modes, the uncertainties in the galaxy bias from the comparison of the
clustering along the transverse and the line of sight directions
because the bias affects only the density growth not the velocity
growth along the line of sight. Combining these different probes on
the galaxy bias would lead to a better handle on the determination of the
galaxy bias and consequently a better measurement of the other
cosmological parameters which suffer from the degeneracies with the
galaxy bias. For instance, the resultant improvement in estimating the
growth rate results in the better estimation of the equation of
state parameter $w$ \cite{gata}, and hence the degeneracy between
$\nmsum$ and $w$ can be broken to improve the neutrino mass
constraint.
\\

\begin{figure}[t]
\begin{center}
\includegraphics[clip,keepaspectratio=true,width=0.9
  \textwidth]{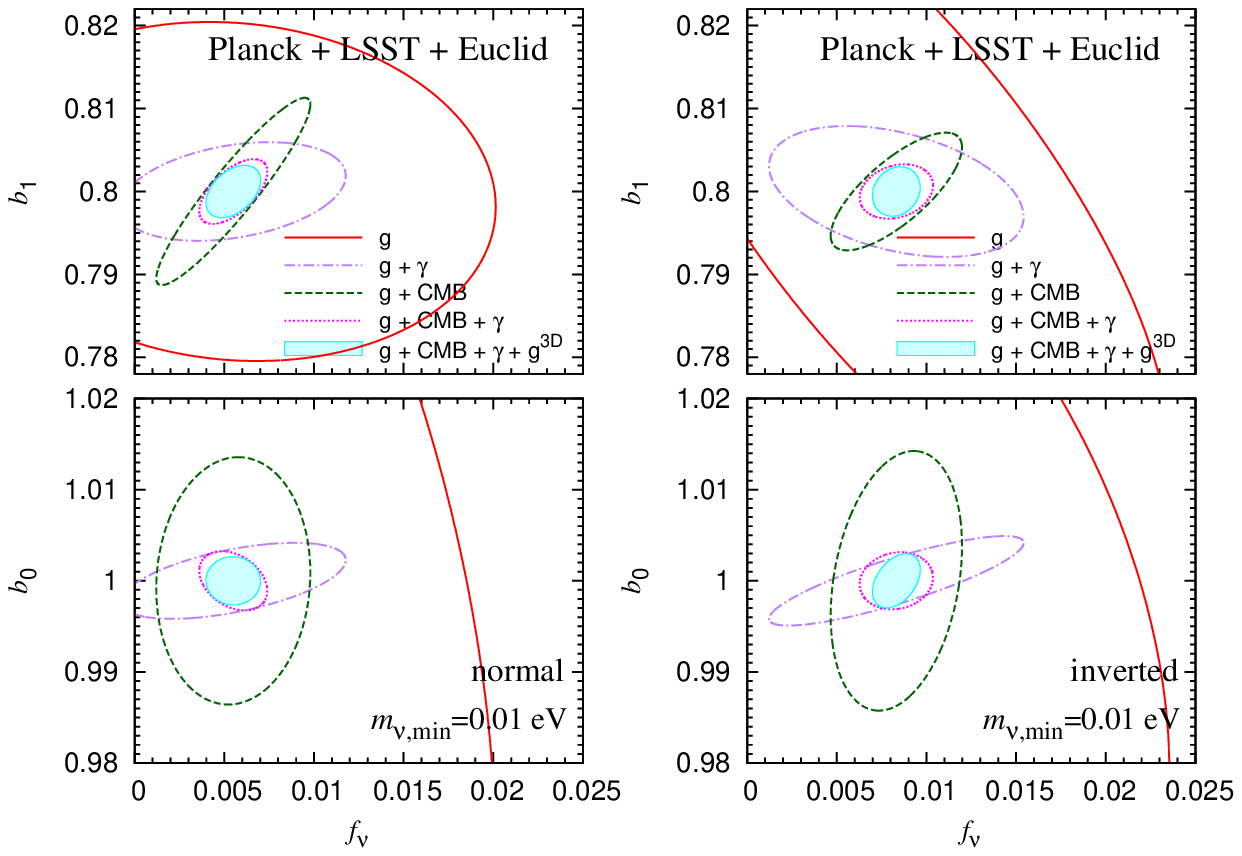}
\end{center}
\caption{
  The degeneracies with the galaxy bias parameters ($b_0$,$b1$). 
  The contours show the the projected 1-$\sigma$ CL areas on the $f_\nu$-$b_0$
  (Bottom) and $f_\nu$-$b_1$ (Top) planes for the normal (Left) and
  inverted (Right) mass hierarchy scenarios with the lightest
  neutrino mass $m_{\nu,{\rm min}}=0.01$ eV.
  Each line and filled area represent the constraints with the
  different combinations of the observables from Planck (CMB), LSST (${\rm g}$ and
  $\gamma$) and Euclid (${\rm g}^{\rm 3D}$);  1) the galaxy clustering alone
  (red), 2) the galaxy clustering and the galaxy weak lensing (purple),
  3) the galaxy
  clustering and CMB (green), 4) the galaxy clustering, the galaxy weak lensing
  and the CMB (magenta), 5) the galaxy clustering, the galaxy
  lensing, CMB and
  the three-dimensional galaxy power spectrum (cyan).
}
\label{fig:bias}
\end{figure}

In Figure~\ref{fig:bias}, we show the projected 1-$\sigma$ CL contours
on the $f_\nu$-$b_0$ and $f_\nu$-$b_1$ planes, i.e. the degeneracies
between the effects of non-relativistic neutrinos and the galaxy
biasing parameters. Although the galaxy biasing parameters have strong
degeneracies with $f_\nu$, the combination of the multiple observables
with the different dependence on the bias parameters indeed helps breaking
the degeneracies with the galaxy bias. The observables such as the CMB
and the galaxy weak lensing ($\gamma$) which are not directly affected
by the galaxy bias are also useful for the improvement on the bias
estimations because they induce the cross
correlations with the observables directly affected by the bias.

Figure~\ref{fig:hierarchy} shows the predicted uncertainties in 
$\sum m_{\nu}$ as a function of the lightest neutrino mass 
$m_{\nu,{\rm min}}$. We should note that the mass splittings are fixed to be
consistent with the oscillation data whose values are given by
Eq.~(\ref{eq:dmv}). Our future cosmological observables we have been
discussing can probe the sum of neutrino masses and still would not be
sensitive enough to differentiate each neutrino mass. Even though the
neutrino effects on the cosmological observables become bigger and
hence a bigger total neutrino mass results in the smaller parameter
uncertainties, the distinction between the normal and inverted mass
patterns becomes harder for a sufficiently large $m_{\nu,{\rm min}}$ leading
to the quasi-degenerate mass spectra $m_1 \sim m_2\sim m_3$. On the
other hand, the smaller neutrino mass leads to the smaller effects on
the matter power and consequently results in the bigger uncertainties in the
parameter estimations. We fortunately find that the neutrino mass
splitting values provided by the current oscillation data give a big
enough sum of the neutrino masses, even with the lightest neutrino
species being massless, for the forthcoming cosmological experiments
to be capable of distinguishing between the normal and inverted mass patterns
with the predicted one-sigma uncertainties taken into account.

Our analysis can claim that the measurement of $\sum m_{\nu}$ from the
combination of the future cosmological observations can distinguish
between the normal and inverted mass hierarchy scenarios for 
$m_{\nu,{\rm min}} \lesssim 0.005$ eV without the overlap in the error bars.
The relative difference between the normal and inverted mass spectrum
becomes smaller for $m_{\nu,{\rm min}} \gtrsim 0.005$ eV with the
overlapped error bars in $\sum m_{\nu}$, and it would become harder to
cosmologically distinguish the two where the mass spectra would
eventually be rather interpreted, for a sufficiently large
$m_{\nu,{\rm min}}$, as the quasi-degenerate mass spectrum
$m_{\nu_1}\sim m_{\nu_2}\sim m_{\nu_3}$ within the predicted
uncertainties. The constraint on the total neutrino mass saturates for
$m_{\nu,{\rm min}} \lesssim 0.001$ eV where the contribution of the
lightest neutrino mass to the total mass becomes negligible.

We can also obtain further information on the neutrino properties by
combining the cosmological data with those from other neutrino
experiments such as the neutrino double beta decay whose observables have
the different dependence on the minimal neutrino mass or neutrino mass
structure \cite{pdg}. For a common value of $m_{\nu,{\rm min}}$,
$\sum m_{\nu}$ is bigger for the inverted mass spectrum than for the
normal mass spectrum. The neutrino effects on the large-scale
structures or the matter power spectrum hence
generally appear more clearly for the inverted mass hierarchy, which
offers tighter constraints, in particular, on the density parameter of
cold dark matter $\Omega_{\rm c}h^2$, the fraction of the massive
neutrinos $f_\nu$, and the effective number of the radiative
components $\Neff$.  We also note that the inverted (as well as the
normal) mass hierarchy scenario can be excluded or inconsistent with
the cosmological observations if we do not observe the effects of the
massive neutrinos for $\sum m_{\nu}$ below the values given by the
lower boundary of the (blue (red) shaded) $1$-$\sigma$ region in Figure~\ref{fig:hierarchy}.

\begin{figure}[t]
\begin{center}
\includegraphics[bb=50 50 410 215,clip,keepaspectratio=true,width=0.9
  \textwidth]{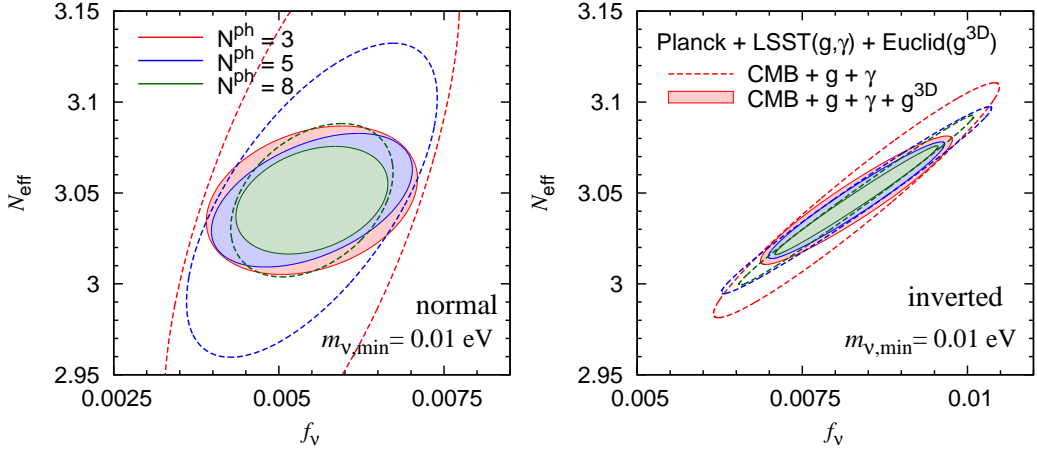}
\end{center}
\caption{
  The comparison of the constraints with different number of
  redshift bins for the photometric redshift survey. The contours
  show the projected 1-$\sigma$ (68\%) CL areas for the normal (Left) and
  inverted (Right) mass hierarchies with the lightest
  neutrino mass $m_{\nu,{\rm min}}=0.01$~eV. For the
  illustrating purpose, we assume the same number of redshift bins
  for the galaxy clustering and the galaxy weak lensing surveys, 
  i.e. $N^{\rm ph}=N_{\rm g}^{\rm ph}=N_{\gamma}^{\rm ph}$. The
  dashed-curves represent the constraints from the CMB, the galaxy
  clustering and the galaxy weak lensing observables, whereas the
  filled-areas represent the constraints from the CMB, the galaxy
  clustering, the galaxy weak lensing observables and the
  3-dimensional galaxy power spectrum from the spectroscopic redshift survey. The different colors show the
  different number of the photometric redshift bins; $N^{\rm
    ph}=$3 (red), 5 (blue) and 8 (green), respectively. The number of the
  redshift bins of the spectroscopic galaxy redshift survey is fixed
  to be $N^{\rm sp}=10$ for all cases.}
\label{fig:cont_bins}
\end{figure}

\begin{figure}[t]
\begin{center}
\includegraphics[bb=50 50 410 215,clip,keepaspectratio=true,width=0.9
  \textwidth]{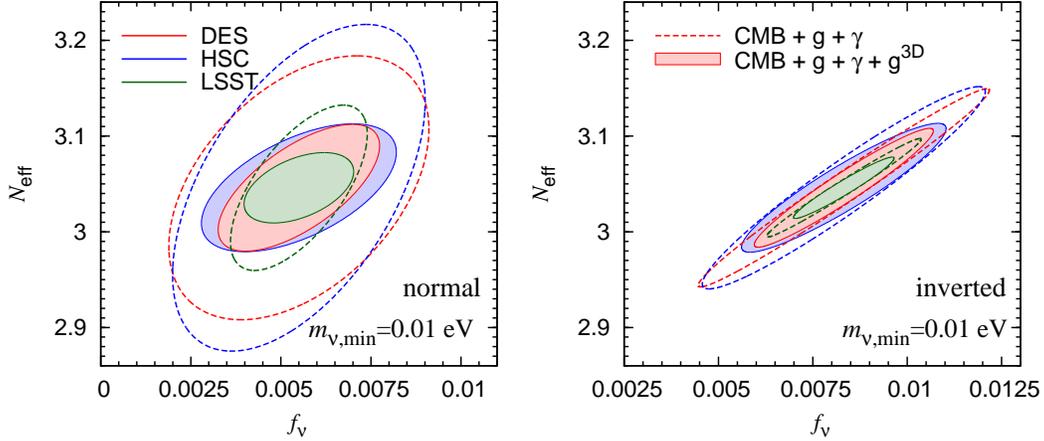}
\end{center}
\caption{
  The comparison among the different photometric redshift
  surveys. The contours show the projected 1-$\sigma$ (68\%) CL areas for
  the normal (Left) and inverted (Right) mass hierarchies, and we here use the fiducial model with the lightest neutrino
  mass of $m_{\nu,{\rm min}}=0.01$~eV. The dashed-curves represent the
  constraints from the combination of the CMB, the galaxy clustering
  and the galaxy weak lensing observables, whereas the filled-areas
  represent the constraints from the CMB, the galaxy clustering, the
  galaxy weak lensing and the 3-dimensional galaxy power spectrum. The
  different colors represent the different photometric redshift
  surveys; DES (red), HSC (blue) and LSST (green), respectively, and we
  assume Planck for CMB and Euclid for the spectroscopic galaxy
  redshift survey for all cases.}
\label{fig:cont_survey}
\end{figure}


\section{Discussion and conclusion}
\label{sec:discuss}

Although we so far presented our results with the concrete fiducial
experiment setups, we now study how the neutrino parameter estimations
would change if we change the survey specifications.  For an
illustration purpose, we here compare the different photometric
redshift survey specifications, firstly by changing the redshift bin
numbers of the fiducial LSST survey and secondly by considering,
instead of the future LSST survey, the on-going photometric redshift
surveys: DES and HSC.

While our fiducial survey model assumed the number of redshift bins
for the photometric redshift survey to be $N^{\rm ph}=5$, the number
of redshift bins is one of the key components in the survey strategy
affecting the constraints on the cosmological parameters. We hence
here briefly discuss how the parameter estimation can be affected by
varying the number of the photometric redshift bins $N^{\rm ph}$. We
assume the same number of the redshift bins for the galaxy clustering
and the galaxy weak lensing observables, $N^{\rm ph}=N^{\rm ph}_{\rm g}=N^{\rm ph}_{\gamma}$, for
simplicity. Figure~\ref{fig:cont_bins} shows the
results on the $f_\nu$-$N_{\rm eff}$ plane for the different number of
the photometric redshift bins, $N^{\rm ph}=$3, 5 and 8. We compare two cases for each number of
$N^{\rm ph}$; one is the combination of the Planck and the LSST
surveys, and the other is the combination of the Planck, the LSST and
the Euclid surveys. The number of redshift bins for the spectroscopic
survey is fixed to be $N^{\rm sp}=10$ and the lightest
neutrino mass $m_{\nu,{\rm min}}=0.01$~eV was used for all cases.

The normal mass hierarchy scenario benefits more from the increase in
the number of redshift bins,
whereas the results for the inverted hierarchy are saturated around
$N^{\rm ph}=5$-8. This is partly because the cosmological observables
are sensitive to the total neutrino mass and the normal mass
spectrum leads to the smaller sum of the
neutrino masses than the inverted one for a common value of the lightest neutrino mass.

\begin{figure}[t]
\begin{center}
\includegraphics[clip,keepaspectratio=true,width=0.9
  \textwidth]{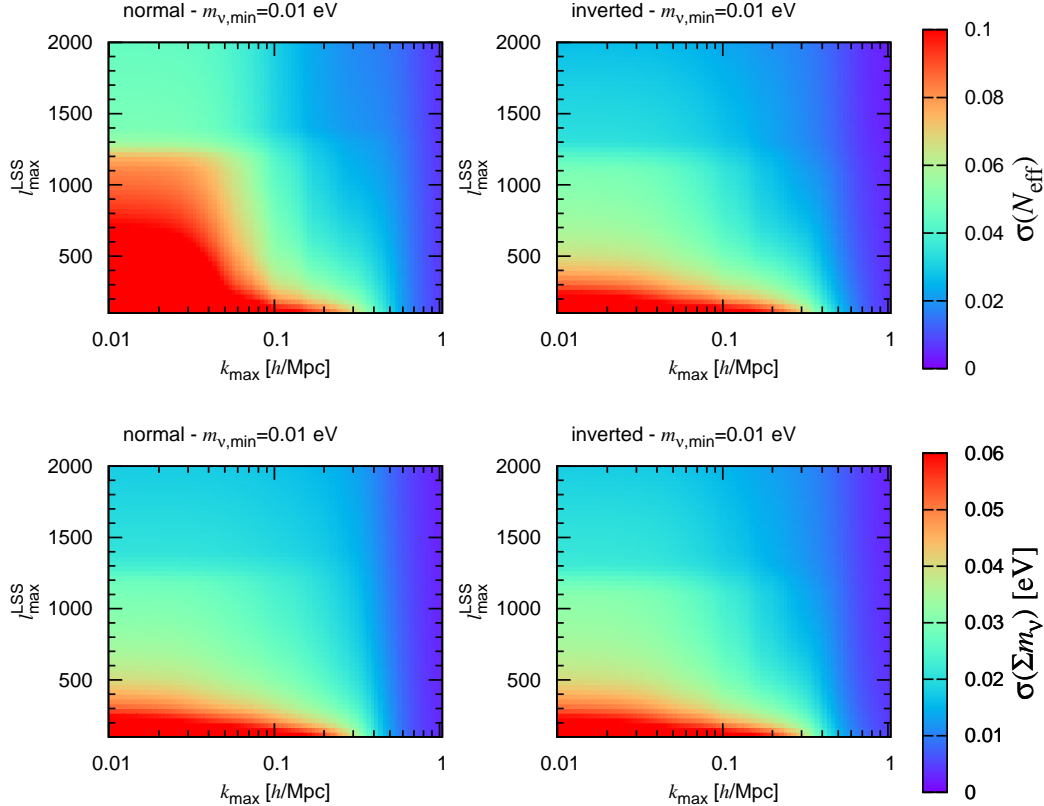}
\end{center}
\caption{
  The dependence of the projected 1$\sigma$ error of the
  neutrino parameter on the maximum wave-number $\kmax$ for the
  spectroscopic galaxy redshift survey, and the maximum multipole
  moment $\lmax^{\rm LSS}$ for the photometric redshift surveys. The
  color bars represent the projected 1-$\sigma$ constraint on the
  effective number $\sigma(N_{\rm eff})$ (Top) and the total neutrino
  mass $\sigma(\sum m_\nu)$ (Bottom) for the normal (Left) and
  inverted (Left) mass hierarchies with the lightest neutrino mass of
  $m_{\nu,{\rm min}}=0.01$ eV. We put the constraint on each parameter
  from the combination of the Planck, LSST and Euclid surveys (CMB + g
  + $\gamma$ + g$^{\rm 3D}$) and the maximum multipole moment for the
  CMB observables is fixed to be $\ell_{\rm max}=3000$. }
\label{fig:small}
\end{figure}

The next comparison we will make here is for the different photometric
redshift surveys to see their effects on the the neutrino property
estimations.  Even though we so far focused on the photometric
redshift survey by the LSST as an ultimate survey in the next decade,
the forecasts from the on-going experiments would be of great interest
and we here investigate how the neutrino properties can be constrained
by the DES and HSC surveys. The survey parameters for the DES and HSC are shown in Table~\ref{tb:survey_phz}.
Figure~\ref{fig:cont_survey} shows the comparison on the
$f_\nu$-$\Neff$ plane for both the normal and inverted mass
hierarchies with $m_{\nu,{\rm min}}=0.01$~eV. Analogously to
Figure~\ref{fig:cont_bins}, we show two kinds of constraints for the
combination of the observables; one is for the combination of the
Planck and the LSST surveys, and the other is for the combination of
the Planck and the LSST surveys plus the spectroscopic redshift survey by the Euclid. 
The constraints on the neutrino parameters from the DES and
HSC surveys are comparable, whereas those from the LSST survey are
much tighter. The differences of constraints between the DES and HSC
surveys come from the depth of the observing redshift range and the
survey area, and the DES survey is slightly superior to the HSC survey
for the constraints on the neutrino properties in this case. For the
total neutrino mass, the DES and HSC surveys can potentially achieve
$\sigma(\sum m_\nu)\sim$0.03~eV in our example, even though the LSST can 
reduce the neutrino mass uncertainties by as much as 50\% for the
normal hierarchy and even more for the inverted hierarchy by a factor
2.

Before concluding our discussions, let us lastly point out that we
conservatively restricted our analysis to the linear regime with the
corresponding choice for $\ell_{\rm max}$ and $k_{\rm max}$. The inclusion
of more modes covering the bigger $\ell_{\rm max}$ and $k_{\rm max}$ would
be of particular interest for the neutrinos sensitive to the small
scale structures.  We conservatively chose the relatively small values for $\ell_{\rm
  max}$ and $k_{\rm max}$ to avoid taking account of the non-Gaussian error estimation 
at the non-linear scales, even though we could have chosen larger
values for higher redshift bins as discussed in Sec.~\ref{sec:fisher}.

For an illustration purpose, we try extending our linear analysis to the
smaller scales to see how the constraints are influenced under the
assumption of the Gaussian error estimation, though the results may
over/under predict the constraints due to the lack of the non-Gaussian
error estimation \cite{Sato:2013,Kayo:2013a}. The estimation of the
matter power spectrum at the non-linear scales may also affect the
constraints of the neutrino properties and we refer the readers to,
for instance, Ref. \cite{sph,bir} for the more appropriate
treatments including the non-linearities in existence of the light
neutrino species. We show the $k_{\rm max}$ and $\ell_{\rm max}$
dependence of the uncertainties in the effective number of neutrinos
and the total neutrino mass parameters in Figure~\ref{fig:small} which
assumes the data from the CMB, galaxy clustering, galaxy weak lensing
observables and the 3-dimensional galaxy power spectrum with the Planck,
LSST and Euclid. We use the same value $\ell_{\rm max}^{\rm LSS}$ for
the maximum multipole of the galaxy clustering and the galaxy weak
lensing observables for comparison purposes, and the maximum multipole
of the CMB observables is fixed to be $\ell_{\rm max}^{\rm CMB}=3000$. 
The fiducial model is the normal or inverted mass
hierarchy scenario with the lightest neutrino mass 
$m_{\nu,{\rm min}}=0.01$ eV.

The constraint on $N_{\rm eff}$ mainly comes from the large-scales via
the shift in the turn-over scale of the matter power spectrum, but the
information from the small-scales helps to break the degeneracy with
other cosmological parameters.  Compared with the inverted mass
hierarchy, the normal mass hierarchy leads to the smaller total
neutrino mass, and consequently the degeneracies with other parameters
are stronger.
For the constraints on $\sum m_{\rm \nu}$, the difference between the
normal and inverted mass hierarchies is not as large as that on
$N_{\rm eff}$, even though the saturation scale of $k_{\rm max}$ for the
normal mass hierarchy is slightly larger than the inverted mass
hierarchy. 
We note that the uncertainty for $\sum m_{\nu}$ saturates at the order
of $\sigma(\Omega_{\rm m} h^2)$ because of the degeneracy between $\sum
m_{\nu}$ and $\Omega_{\rm m} h^2$. 
\\

We studied how much the constraints on the neutrino properties are
improved by combining the various observables from the next-generation
cosmological surveys with the cross-correlations taken into account
among those observables. We in particular studied, for the fiducial
neutrino models, the normal and inverted mass hierarchy scenarios
consisting of three neutrino species where the neutrino mass
splittings are fixed to those values obtained by the neutrino
oscillation measurements and the only free neutrino mass parameter
hence is the minimal neutrino mass. We also studied how the
cosmological parameter errors are affected by the different choices of
a minimal neutrino mass value. For the reference cosmological surveys, 
the Planck CMB, LSST photometric redshift survey and Euclid 
spectroscopic redshift survey are assumed, where the forecasted
uncertainties in the neutrinos were found to be in the range
$\sigma(\sum m_{\nu})\sim 0.015$-0.025 eV and $\sigma(N_{\rm eff})\sim
0.03$. We also estimated the projected cosmological errors assuming
the on-going photometric galaxy surveys such as the DES and HSC to
compare with the LSST, and we found the neutrino mass uncertainties
increase by as much as a factor 2.

The merits of adding the spectroscopic redshift survey information
with the redshift space distortion to the two dimensional observables
including the CMB lensing and galaxy weak lensing survey were
emphasized to break the degeneracies among the parameters, such as
those between the neutrino and dark energy parameters. A better handle
on the galaxy bias was argued to be responsible for the improvement in
the parameter estimation, and, in particular, the improvement for the
neutrino mass estimation amounts to $20-40$\% depending on the
neutrino mass hierarchical structure.  The power of the future
cosmological observables to distinguish between the normal and
inverted neutrino mass hierarchy scenarios was also quantified. We
found the measurement of the total neutrino mass can well distinguish
between the normal and inverted mass hierarchy scenarios for
$m_{\nu,{\rm min}} \lesssim 0.005$ eV. We restricted our analysis to
the linear regime and extending our analysis to a smaller scale in
principle can potentially reduce the projected errors. The outcome
would however heavily depend on the treatment of non-linearities, and
the more proper treatment including for instance the non-Gaussian
covariance would be necessary to probe those smaller scales.

Finally, we should note the accuracy of the Fisher matrix analysis for the constraint on neutrino masses. Especially for small neutrino masses, the constraint by the Fisher matrix analysis might not be completely accurate because the assumption of the Gaussian posterior could be no longer valid, e.g., \cite{Perotto:2006,Galli:2010,hall12,Audren:2013}. The potential quantitative differences between the Fisher matrix and the Monte Carlo Markov Chain (MCMC) approaches for constraining the neutrinos from the CMB observation are presented, for instance, in \cite{Perotto:2006,hall12}, and we leave the the analysis with the MCMC approach and its comparison with that using the Fisher analysis for our future work.

The neutrino properties in combination of other experiments besides the
cosmological observables such as the accelerator neutrinos and
neutrinoless double beta decay experiments would certainly deserve the further
studies as a unique window to the physics beyond the Standard Model.


\section*{Acknowledgement}

This work was supported in part by the Grant-in-Aid for the Global COE
program and for Scientific Research from the JSPS.
We thank Kobayashi-Maskawa Institute for the Origin of
Particles and the Universe; Nagoya University for providing computing
resources in conducting the research reported in this paper, 
the Aspen Center for Physics and the Kavli Institute for Theoretical Physics
China where part of this work was conducted.


\providecommand{\href}[2]{#2}\begingroup\raggedright

\end{document}